\documentclass[11pt]{article}
\usepackage[utf8]{inputenc}
\usepackage{jheppub}
\usepackage{amsmath,amssymb,amsfonts,graphicx,slashed,color,amsthm,mathtools,upgreek, enumerate, tensor, scalerel, subfig}
\usepackage{arydshln}
\usepackage{bbold}
\usepackage[dvipsnames]{xcolor}
\allowdisplaybreaks

\title{Quantum Error Correction in the Black Hole Interior}
\author[a,b]{Vijay Balasubramanian}
\author[c]{\!, Arjun Kar}
\author[a]{\!, Cathy Li}
\author[d]{\!, Onkar Parrikar} 

\affiliation[\,a]{David Rittenhouse Laboratory, University of Pennsylvania,\\
209 S. 33rd Street, Philadelphia PA 19104, USA.}
\affiliation[\,b]{Theoretische Natuurkunde, Vrije Universiteit Brussel (VUB), and \\ International Solvay Institutes, Pleinlaan 2, B-1050 Brussels, Belgium.}
\affiliation[\,c]{Department of Physics and Astronomy, University of British Columbia, \\
6224 Agricultural Road, Vancouver, BC V6T 1Z1, Canada.}
\affiliation[\,d]{Department of Theoretical Physics,
Tata Institute for Fundamental Research,\\ Mumbai 400005, India.}

\newcommand{\beq}{\begin{equation}}
\newcommand{\eeq}{\end{equation}}
\newcommand{\beqn}{\begin{eqnarray}}
\newcommand{\eeqn}{\end{eqnarray}}

\newcommand{\cE}{\mathcal{E}}
\newcommand{\en}{\text{env}}
\newcommand{\cN}{\mathcal{N}}
\newcommand{\ph}{\text{phys}}
\newcommand{\co}{\text{code}}
\newcommand{\re}{\text{ref}}
\newcommand{\cH}{\mathcal{H}}
\newcommand{\cO}{\mathcal{O}}
\newcommand{\inte}{\text{int}}
\newcommand{\ext}{\text{ext}}
\newcommand{\bra}[1]{\langle #1 \vert}
\newcommand{\ket}[1]{\vert #1 \rangle}
\newcommand{\inner}[2]{\langle #1 \vert #2 \rangle}

\DeclareMathOperator{\Tr}{Tr}

\newcommand{\cR}{\mathcal{R}}
\newcommand{\cS}{\mathcal{S}}
\newcommand{\fr}{\mathfrak{R}}

\newcommand{\Abar}{{\overline{A}}}

\begin{document}

\abstract{We study the quantum error correction properties of the black hole interior in a toy model for an evaporating black hole: Jackiw-Teitelboim gravity entangled with a non-gravitational bath. 
After the Page time, the black hole interior degrees of freedom  in this system are encoded in the bath Hilbert space.
We use the gravitational path integral to show that the interior density matrix
is correctable against the action of quantum operations on the bath which (i) do not have prior access to details of the black hole microstates, and (ii) do not have a large, negative coherent information with respect to the maximally mixed state on the bath, with the lower bound controlled by the black hole entropy and code subspace dimension. 
Thus, the encoding of the black hole interior in the radiation is robust against generic, low-rank quantum operations. For erasure errors, gravity comes within an $O(1)$ distance of saturating the Singleton bound on the tolerance of error correcting codes.
For typical errors in the bath to corrupt the interior, they must have a rank that is a large multiple of the bath Hilbert space dimension, with the precise coefficient set by the black hole entropy and code subspace dimension.
}

\maketitle

\parskip=12pt

\section{Introduction}

In the AdS/CFT correspondence, where a precise non-perturbative definition of quantum gravity exists \cite{Maldacena:1997re,Witten:1998qj}, one of the most basic objects  is the bulk-to-boundary map $V$ in the large $N$ and large t'Hooft coupling limit \cite{Balasubramanian:1998sn,Banks:1998dd,Balasubramanian:1999ri,Hamilton:2006az}. In this limit, the bulk description is well approximated by semi-classical supergravity, and the bulk-to-boundary map $V$ embeds  bulk semi-classical states into the Hilbert space $\mathcal{H}_\text{CFT}$ of the dual, strongly coupled, quantum mechanical theory. The bulk semi-classical states in question are essentially comprised of effective field theory (EFT) excitations of bounded energy density around some chosen gravitational background $g$, which we can arrange in the form of a bulk EFT Hilbert space $\mathcal{H}_\text{bulk EFT}^g$. Two decades of work have explored aspects of this embedding of the bulk gravity EFT in the boundary CFT.   One of the most mysterious aspects of the map is the emergence of locality in the bulk low-energy theory, especially in regions causally hidden behind horizons.

Some fundamental properties of the map $V$, such as the redundancy of bulk operator representations and radial locality, are most naturally understood by realizing that the embedding
\begin{equation}
    V: \mathcal{H}_{\text{bulk EFT}}^g \to \mathcal{H}_\text{CFT} ,
\end{equation}
has the structure of a quantum error correcting code \cite{Verlinde:2012cy,Almheiri:2014lwa,Pastawski:2015qua}. In the quantum error correction (QEC) language, states in the bulk effective field theory around the gravitational background $g$ are often referred to as the code subspace $\mathcal{H}_\co \subset \mathcal{H}_\text{CFT}$ \cite{Harlow:2016vwg}.
The code subspace is intuitively the subspace of states in the full Hilbert space $\mathcal{H}_\text{CFT}$ upon which a bulk local EFT operator may be represented, and states outside the code subspace are interpreted as having lost access to the operator either due to absorption by a large black hole or a more fundamental breakdown of the emergent background metric $g$ \cite{Harlow:2018fse}.  Quantum error correction in this holographic context means that the representation on a boundary subregion $A$ of bulk operators within a corresponding bulk subregion enclosed between $A$  and the quantum extremal surface (QES) for $A$ -- called the ``entanglement wedge" $\mathcal{W}(A)$  -- is robust against the erasure of the complementary subregion $\Abar$ \cite{Dong:2016eik}. Equivalently, the state of bulk quantum fields in the region $\mathcal{W}(A)$ is ``encoded'' in the boundary subregion $A$, and is protected from any and all ``errors'', i.e., operator actions, on $\Abar$. Conversely operators in the entanglement wedge $\mathcal{W}(A)$ can be reconstructed entirely from data in $A$.  This modernization of the bulk-to-boundary map is referred to as entanglement wedge reconstruction (EWR).

\begin{figure}[t]
    \centering
    \includegraphics[height=5cm]{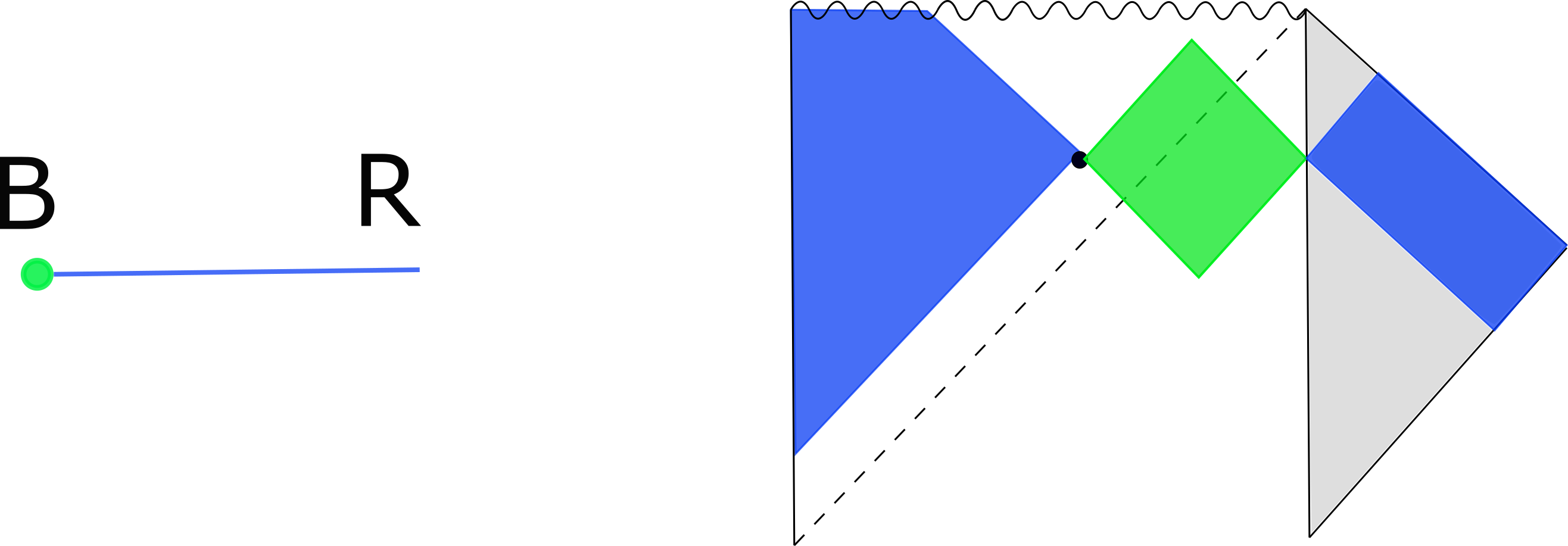}
    \caption{\small{
    A cartoon of an evaporating black hole. \textbf{(Left)} The UV description consists of a holographic quantum mechanical system $B$ (green dot) in a highly excited state coupled to a bath $R$ (blue line). \textbf{(Right)} The dual semi-classical description is that of an evaporating black hole coupled to a bath. Past the Page time, there is a new Quantum Extremal Surface (black dot) relevant for the entropy of the bath, and the region in the black hole interior enclosed by it to its left is called the ``island''. 
    }
    }
    \label{fig:SC}
\end{figure}

The EWR paradigm has provided a powerful framework for bringing tools from quantum information theory to bear on problems in quantum gravity. For instance, EWR played an important role in recent progress on the black hole information problem \cite{Almheiri:2019psf,Penington:2019npb,Almheiri:2019qdq,Penington:2019kki,Chen:2019iro}. These papers studied an asymptotically AdS$_2$ evaporating black hole coupled to a non-gravitational bath, where the bath serves as a reservoir to collect the Hawking radiation (Fig.~\ref{fig:SC}). In this context, if one computes the von Neumann entropy of the bath past the Page time, the semi-classical description involves a new quantum extremal surface that develops in the black hole interior and cuts off the naive growth of the entropy, reproducing the expected Page curve. Crucially, the EWR paradigm in this context implies that the region in the black hole interior enclosed by this new QES -- the so called ``island'' -- is encoded in the bath, and thus bulk operators in the island can be represented as operators on the bath. Indeed, it was shown in \cite{Penington:2019kki} that a general tool from the theory of quantum error correction called the Petz map can be used to give an explicit reconstruction of operators in the black hole interior.  

This is a surprising and radical claim -- the bath need not be a conventional, strongly coupled holographic quantum field theory, and could be a free field theory, for instance. Yet, it can  encode semi-classical supergravity degrees of freedom in the same sense as holographic duality, i.e., a portion of the bath is dual to, or an equivalent description of, the gravity degrees of freedom in a distant part of space.  In an evaporating black hole, this would  appear to be at odds with the semi-classical causal structure of the black hole spacetime, as the encoded island lies behind the horizon.  However, this encoding should be fundamentally different from the standard bulk-boundary map in holographic dualities: we expect it to be exponentially complex in the black hole entropy, i.e, implementing this encoding with a quantum computer should be exponentially complex in $N$, explaining why simple observables give the appearance of information loss in Hawking radiation.  By contrast, the standard encodings of EFT excitations in regions of AdS space that are causally accessible from the boundary only have polynomial complexity since the reconstruction just involves solving equations of motion \cite{Balasubramanian:1998sn,Balasubramanian:1999ri,Hamilton:2006az}.  Note that for entanglement wedges of a strongly coupled theory which include a ``Python's Lunch'', which we will discuss below, the encoding map is also expected to be highly complex since the main source of reconstructive power is entanglement between gravitational degrees of freedom.

Recently, progress was made toward a more detailed understanding of the quantum error correcting structure of the code in the Hawking radiation for the interior of the black hole \cite{Kim:2020cds}. In this paper, Kim, Preskill and Tang (KPT) suggested that the bulk degrees of freedom in the black hole interior are encoded in the radiation with robust QEC properties -- more so than the usual QEC familiar in encoding of ``causal wedges'', i.e., regions causally connected to the spacetime boundary in AdS/CFT. In particular, they found (in a model for black hole evaporation) that despite the fact that the ``island'' region in the black hole interior  is encoded in the early Hawking radiation, this encoding is protected in the QEC sense from all ``low-complexity'' and ``low-rank'' errors on the early radiation as long as the remaining black hole is macroscopic.  Here an ``error'' is any quantum operation that acts on the radiation. This is very different from the standard statement of QEC in AdS/CFT where the bulk entanglement wedge $\mathcal{W}(A)$ is protected from errors in $\Abar$, but not in general from errors on $A$ itself.  KPT further argued that this robust error correction  has an important bearing on the apparent tension between islands and the semi-classical causal structure of the black hole -- all low complexity quantum operations on the bath commute with the representations in the bath of bulk operators in the black hole interior. Thus, in order to affect the black hole interior explicitly, an observer in the bath will have to perform an exponentially complex operation. Crucial to the analysis of KPT was an assumption that the reduced state on the bath/radiation is \emph{pseudorandom}, i.e., cannot be reliably distinguished from the maximally mixed state (more generally, the thermal ensemble) with a polynomial complexity quantum operation; complexity theory enters their arguments through this assumption. 

The main purpose of our work is to study these novel  quantum error correction properties of the encoding of the black hole interior in Hawking radiation.  Instead of using the KPT pseudorandomness assumption, we will  directly study the QEC properties in a toy model for an evaporating black hole in Jackiw-Teitelboim (JT) gravity using the Euclidean gravity path integral. We will show that in this setting the bulk density matrix in the interior is approximately recoverable from the action of quantum operations on the bath which (i) do not have prior access to the details of the black hole microstates and their overlaps, and (ii) do not have a large, negative ``coherent information'' -- an information theoretic measure of the amount of noise a quantum operation generates -- with respect to the maximally mixed state on the bath (which we may interpret as the ``semi-classical state" of the radiation, i.e., the analog of Hawking's thermal state). The lower bound on the coherent information is controlled by the black hole entropy and code subspace dimension. This implies that the encoding of the black hole interior in the radiation is robust against generic, low-rank quantum operations on the bath which do not have access to details of the black hole microstate, thus providing a gravitational perspective on the results of KPT. Furthermore, for erasure errors on the bath, we show that gravity comes within an $O(1)$ distance of saturating the quantum Singleton bound on the error tolerance of quantum error correcting codes.  We also show that typical errors on the bath must have a large rank compared to the product of the black hole Hilbert space and bath Hilbert space dimensions in order to affect the interior.
We  comment on the interpretation of our coherent information criterion in light of an unusual non-isometry property of the interior encoding map that we describe.

While our analysis here is limited to the specific example of the black hole interior, we propose that even for a more general entanglement wedge $W(A)$ dual to a CFT subregion $A$, \emph{the encoding of semi-classical degrees of freedom in the python's lunch} \cite{Brown:2019rox, Engelhardt:2021mue, Engelhardt:2021qjs} \emph{(i.e., the part hidden behind a non-minimal QES) should be robust against generic, low-rank errors on the boundary subregion $A$.} By ``generic'', we mean quantum operations which do not have prior access to the details of the encoding map. Note that such robust error correction properties certainly do not hold in the causal wedge. Indeed, after the Page time the black hole interior is in the python's lunch because it is hidden behind a non-minimal QES, namely the empty surface. Finally, we discuss  implications of our results for the consistency of the island program  for black hole information recovery \cite{Almheiri:2019psf,Penington:2019npb,Almheiri:2019qdq,Penington:2019kki,Chen:2019iro} with the semi-classical causal structure of evaporating black hole spacetimes.

Four sections follow.
In Sec.~\ref{sec:prelim}, we review necessary ideas from quantum error correction as well as the toy model of black hole evaporation which serves as our primary example.
In Sec.~\ref{sec:qec-pssy}, we apply  QEC tools to the toy model and, using the gravitational path integral, derive our criterion for correctable errors on the black hole interior code subspace. Euclidean wormholes play a key role in the action of errors on the encoded state which are not correctable.  We also discuss the non-isometric nature of the interior encoding. In Sec.~\ref{sec:ErasuresTypical} we present a detailed analysis of erasure errors and typical/random, fixed-rank errors.
We conclude with a discussion of generalization to other entanglement wedges and implications for the semi-classical causal structure of the evaporating black hole spacetime in Sec.~\ref{sec:disc}.

\section{Preliminaries}\label{sec:prelim}
In this section, we review background material on quantum error correction, following \cite{NielsenChuang}, and the toy model of black hole evaporation \cite{Penington:2019kki} which we will use in later sections. The material in this section is not new, but is included here to keep the article somewhat self-contained. 

\subsection{Conditions for quantum error correction}\label{sec:QECreview}

Let $\cH_{\co}$ denote the code subspace, and let $\{|i\rangle\}_{i=1}^d$ be a basis of states for $\cH_{\co}$. Furthermore, let $\mathcal{H}_{\text{phys}}$ be a larger Hilbert space in which we wish to encode $\mathcal{H}_{\text{code}}$, via the isometric encoding map:
\beq
V: \mathcal{H}_{\text{code}} \to \mathcal{H}_{\text{phys}} , \quad V\ket{i}_\co = \ket{\Psi_i}_\ph,\quad V^{\dagger}V=\mathbb{1}_\co.
\eeq
Thus $VV^\dagger$ projects the physical Hilbert onto encoding of the code subspace in the physical degrees of freedom. The general theory of quantum error correction deals with recovering the encoded state from the physical system after the action of quantum operations, which can potentially create ``errors'' in the encoding.  Errors in the physical system can be modelled by the action of a quantum channel $\mathcal{E}$ which is a linear, completely positive\footnote{All operations of the form $\cE(\rho) = \sum_m E_m \rho E_m^\dagger$ are completely positive.    
Complete positivity of $\cE$ means $(\mathbb{1}_E \otimes \cE)(A) \geq 0$ for any positive $A \geq 0$ which acts on the extended system $\cH_E \otimes \cH_\ph$.
For any $\ket{\psi}_{E,\ph}$, we define $\ket{\phi_m}_{E,\ph} = (\mathbb{1}_E \otimes E_m^\dagger) \ket{\psi}_{E,\ph}$.
By $A \geq 0$ we have $\bra{\phi_m} A \ket{\phi_m} \geq 0$.
Summing over $m$, we find the action of $\mathbb{1}_E \otimes \cE$ on $A$:
\[ \sum_m \bra{\phi_m} A \ket{\phi_m} = \sum_m \bra{\psi} (\mathbb{1}_E \otimes E_m) A (\mathbb{1}_E \otimes E_m^\dagger) \ket{\psi} \geq 0 . \]
The middle expression is nothing but $\bra{\psi} (\mathbb{1}_E \otimes \cE)(A) \ket{\psi}$, and we have shown it is non-negative for any $\ket{\psi}$, so we have $(\mathbb{1}_E \otimes \cE)(A) \geq 0$.  These conditions guarantee in our setting that  density matrices on the system plus the environment are mapped to density matrices.
}
and trace preserving map which takes a state (i.e., a density matrix\footnote{In our notation, we will refer to both pure states $\ket{\psi}$ and density matrices $\rho$ as ``states". }) on $\mathcal{H}_{\text{phys}}$ to another state on $\mathcal{H}_{\text{phys}}$.

The action of a general quantum channel can always be expressed in terms of its Kraus operators $\{E_m\}$:
\beq
\cE(\rho) = \sum_m E_m \rho E_m^{\dagger},
\eeq
where requiring $\cE$ to be trace preserving implies 
\beq \label{trC}
\sum_m E_m^{\dagger} E_m = \mathbb{1}.
\eeq
The minimum number of Kraus operators necessary to implement $\cE$ is sometimes called the (Kraus) rank of the channel, and could be regarded as measure of how complex a channel is.\footnote{Obviously, the rank is not related to the circuit complexity of the channel since a channel of rank 1 has a single unitary Kraus operator which can have arbitrary circuit complexity.}
Another particularly useful fact is that the action of any quantum channel $\cE$ on a state $\rho$ can be modeled by a unitary operation on an enlarged Hilbert space that includes an auxiliary environment.  In detail, introduce an environment  $\mathcal{H}_{\en}$ spanned by $\{|e_m\rangle\}$ in some  initial state $|e_0\rangle$. Then act on the enlarged state $\rho\otimes |e_0\rangle\langle e_0| \in \mathcal{H}_{\text{phys}}\otimes \mathcal{H}_{\en}$ with a  unitary operator $U_\cE$, and  trace out the environment to get
\beq\label{SD}
\cE(\rho) = \mathrm{Tr}_{\en}\left[ U_\cE( \rho \otimes |e_0\rangle\langle e_0|_{\en} ) U_\cE^{\dagger} \right],
\eeq
where $E_m = \langle e_m | U_\cE | e_0\rangle$, i.e., $U_\cE|\psi\rangle\otimes |e_0\rangle = \sum_{m}E_m |\psi\rangle \otimes |e_m\rangle$. 
This expression of the channel $\mathcal{E}$ as the trace over an environment after conjugation by a unitary $U_\cE$ will be referred to as the \emph{isometric extension} (sometimes also called the  Stinespring dilation) of $\mathcal{E}$. It is a way of representing a quantum channel as an open system.  

We say that an error $\mathcal{E}$ can be corrected on $\cH_{\co}$ if there exists a recovery channel $\mathcal{R}$ such that
\beq
\mathcal{R}(\mathcal{E}(\rho)) = \rho, 
\eeq
for any density matrix $\rho$ supported on $\mathcal{H}_{\text{code}}$, or more precisely, supported on the image of the code subspace inside the physical Hilbert space under $V$. One of the central results of the theory of quantum error correction is that the channel $\cE$ is exactly correctable if and only if it satisfies the  ``Knill-Laflamme'' conditions \cite{NielsenChuang}:
\beq \label{KLcond0}
P_{\text{code}} E_m^{\dagger}E_n P_{\text{code}} = \alpha_{mn}P_{\text{code}},
\eeq
where $P_{\text{code}} = VV^\dagger$ is the projector onto the code subspace.  
The projection, trace preservation, and positivity properties of the operators on the left hand side of \eqref{KLcond0} imply that $\alpha$ is actually a density matrix on $\cH_\en$.

It is useful to restate this criterion in a different form. We introduce a reference system with a Hilbert space $\cH_\re$ isomorphic to the code Hilbert space and construct the state 
\beqn
|\Psi'\rangle &\equiv& \frac{1}{\sqrt{d}}\sum_{i}|i\rangle_{\re}\otimes U_\cE(|\Psi_i\rangle_{\ph}\otimes |e_0\rangle_{\en}) \nonumber\\
&\equiv& \frac{1}{\sqrt{d}}\sum_{i,m}|i\rangle_{\re}\otimes (E_m|\Psi_i\rangle_{\ph})\otimes |e_m\rangle_{\en} \, ,
\eeqn
where $|\Psi_i\rangle_\ph=V|i\rangle_{\co}$ are the basis states of the code subspace of the physical Hilbert space.
The state $\ket{\Psi'}$ can be thought of as the action of the error channel on a state that is maximally entangled  between the code subspace and an auxiliary reference system, and is engineered so that its entanglement properties are related to the validity of \eqref{KLcond0}.
Specifically, the error $\cE$ is correctable if and only if the reduced state of $\ket{\Psi'}$ on $\cH_\re \otimes \cH_\en$ factorizes as \cite{Schumacher:1996dy}
\beq
\rho'_{\re,\en} = \rho'_{\re}\otimes \rho'_{\en} ,
\eeq
where the reduced density matrices above are defined as $\rho'_{\re,\en} = \Tr_\ph \ket{\Psi'}\bra{\Psi'}$ and so on.
This is called the \emph{decoupling principle} \cite{Preskill:2016htv}.

We can diagnose  decoupling by computing the mutual information:
\beq
I_{\Psi'}(\re:\en) \equiv S_{\Psi'}(\re)+S_{\Psi'}(\en)-S_{\Psi'}(\re\cup \en),
\eeq
where each term on right hand side is the entropy of the reduced density matrix on the indicated factor. This mutual information vanishes if and only if the decoupling principle holds. The decoupling principle also generalizes to approximate quantum error correction \cite{2001quant.ph.12106S}. For instance, we will often be interested in cases where the above mutual information is not strictly zero, but small (i.e., exponentially small in the black hole entropy). In such a situation, the mutual information gives a bound on the decoupling:
\beq
||\rho'_{\re,\en}-\rho'_{\re}\otimes \rho'_{\en}||_1 \leq \sqrt{2 I_{\Psi'}(\re:\en)},
\eeq
where $||A ||_1 = \mathrm{Tr}\,\sqrt{A^{\dagger}A}$ is the trace norm. This in turn implies that the error channel is approximately correctable \cite{Kim:2020cds}, i.e., there exists a recovery channel $\cR$ such that
\beq
\max_{\rho}||\cR(\cE(\rho))-\rho||_1 \leq \epsilon,\;\;\;\epsilon = \left(\frac{1}{2}I_{\Psi'}(\re:\en)\right)^{1/4},
\label{eq:mutual-bound}
\eeq
where the maximization is over all states supported on the code subspace. 

Using the decoupling principle, we can derive a simple bound on the number of maximal errors, i.e., erasures, a code can tolerate while remaining correctable. Let $t-1$ be the maximum number of qubits in the physical Hilbert space on which we can tolerate errors. Let $d$ be the dimension of the code space and let $2^{\cS}$ be the dimension of the physical Hilbert space. Partition the physical Hilbert space into three subsets: $Q_1$ and $Q_2$ with $(t-1)$ qubits each and $Q_3$ with the remaining $(\cS-2t+2)$ qubits. Now consider the state 
\beq
|\chi\rangle = \frac{1}{\sqrt{d}}\sum_i |i\rangle_{\re}\otimes |\Psi_i\rangle_{\ph}.
\eeq
Since both $Q_1$ and $Q_2$ have less than $t$ qubits each, we can, by definition, tolerate errors on these subsystems. For instance, we could bring in $(t-1)$ environment qubits and swap the state on $Q_1$ or $Q_2$ with the environment. For such operations to be correctable, by the decoupling principle there must be no correlations after the swap between the environment and the reference. This in turn means that the initial state cannot have correlations between the qubits in $Q_1$ and the reference, and similarly for $Q_2$. Using this fact, we get
\beq
S(\re) + S(Q_1) = S(\re,Q_1) = S(Q_2Q_3) \leq S(Q_2) + S(Q_3),
\eeq
\beq
S(\re) + S(Q_2) = S(\re,Q_2) = S(Q_1Q_3) \leq S(Q_1) + S(Q_3),
\eeq
where all the entropies refer to the state $|\chi\rangle$, and we have used the subadditivity of entropy in the last step of each inequality. The second equality follows because the entropy of a subsystem of a pure state equals the entropy of its complement. Adding these inequalities  gives
\beq
S(\re)\leq S(Q_3)\;\;\;\Rightarrow 2(t-1) \leq (\cS-\log_2d).
\label{eq:singleton-bound}
\eeq
where we used the fact that the reference state is maximally entangled with the code subspace and so $S(\re) = \log_2{d}$ while the $S(Q_3)$ is bounded from above by the number of qubits in $Q_3$, namely $(\cS-2t+2)$.  This last inequality is known as the \emph{quantum Singleton bound.}

\noindent\textbf{Subsystem QEC condition}:  We will be interested in a setting where the code subspace has a natural tensor factorization $\cH_{\co}=\cH_a\otimes \cH_{a'}$. In the black hole case this will correspond to  factorization of some subset of the bulk quantum field theory degrees of freedom between the interior and the exterior of the black hole. With this additional structure, we can  define a more general version of error correction \cite{2005quant.ph..4189K, Nielsen_2007}, which, following \cite{Harlow:2016vwg}, we will refer to as ``subsystem quantum error correction''. We say that the state on $\cH_{a}$ is recoverable under the error $\cE$ if, for any density matrix $\rho$ supported on $\mathcal{H}_{a}$ and any density matrix $\sigma$ supported on $\cH_{a'}$, there exists a recovery channel $\mathcal{R}$ such that
\beq
\mathcal{R}(\cE(\rho \otimes \sigma)) = \rho \otimes \sigma',
\eeq
for some density matrix $\sigma'$ on $\cH_{a'}$. We again have in mind that the density matrix on the left hand side has been encoded in $\cH_\ph$ by the map $V$ before the action of $\cE$. In analogy with equation \eqref{KLcond0} the channel $\cE$ is correctable on $\cH_a$ if and only if 
\beq \label{KLcond1}
P_{\text{code}} E_m^{\dagger}E_n P_{\text{code}} = \mathbb{1}^{(a)}\otimes B^{(a')}_{mn},
\eeq
where $P_{\text{code}}$ is the projector onto the code subspace, and $B_{mn}$ are some operators with support only on $\cH_{a'}$ \cite{2005quant.ph..4189K, Nielsen_2007} . 

Following \cite{Nielsen_2007}, it is useful to restate this criterion in the form of a decoupling principle. In order to do this, we once again need to introduce a reference system. Let us in fact introduce two reference systems $\fr_a$ and $\fr_{a'}$ which have the same dimensions as $\mathcal{H}_{a}$ and $\mathcal{H}_{a'}$ respectively. For any basis state $|i, i'\rangle_{\co} \equiv |i\rangle_a \otimes |i'\rangle_{a'}$, let $|\Psi_{i,i'}\rangle_{\ph}$ be the image of $\ket{i,i'}_\co$ in $\cH_{\ph}$ under the encoding map $V$: 
\beq
V|i,i'\rangle_{\co} = |\Psi_{i,i'}\rangle_{\ph} .
\eeq
We now introduce a basis for the reference system $\ket{i,i'}_\re = \ket{i}_{\fr_a} \otimes \ket{i'}_{\fr_{a'}}$, and an environment $\cH_{\en}$ supporting isometric extension of the quantum channel $\cE$.  In terms of these quantities we define 
\begin{align} 
\label{psi'}
\ket{\Psi'} & \equiv \frac{1}{\sqrt{d}}\sum_{i,i'} \ket{i,i'}_\re \otimes U_{\cE} (\ket{\Psi_{i,i'}}_\ph \otimes \ket{e_0}_\en ) \\
& = \frac{1}{\sqrt{d}} \sum_{i,i',m} \ket{i,i'}_\re \otimes E_m \ket{\Psi_{i,i'}}_\ph \otimes \ket{e_m}_\en .
\end{align}
We are now in a position to restate the condition for subsystem quantum error correction: an error $\cE$ is correctable on $\cH_a$ if and only if the reduced state $\rho'$ corresponding to $|\Psi'\rangle$ on the subsystem $\cH_\re \otimes \cH_\en$ factorizes as
\beq
\rho'_{\fr_a,\fr_{a'},\en} = \rho'_{\fr_a} \otimes \rho'_{\fr_{a'},\, \en} \, .
\eeq
Equivalently,
\beq \label{SQEC}
I_{\Psi'}(\fr_a: \fr_{a'}\cup \en) \equiv S_{\Psi'}(\fr_a) + S_{\Psi'}(\fr_{a'}\cup \en) - S_{\Psi'}(\fr_a \cup \fr_{a'} \cup\en) = 0.
\eeq
It is particularly useful to restate the QEC conditions in the form of a vanishing mutual information because it distills down the criterion to the computation of specific entanglement entropies. In AdS/CFT, the gravitational path integral can compute entanglement entropies via the replica trick. Thus, we can hope to get a handle on the error correction properties of subsystems in gravity by studying the above mutual information via the gravitational path integral.  Indeed, for the black hole interior, the only way we can currently access the bulk-to-boundary encoding map $V$ is indirectly via the Euclidean path integral.  In Sec.~\ref{sec:qec-pssy}, our primary goal is to compute this mutual information with $\fr_a$ and $\fr_{a'}$ being the reference systems for the interior and exterior code subspaces on an evaporating black hole background after the Page time.

\subsection{PSSY model}
In this section, we review a toy model for an evaporating black hole that was constructed by Penington, Shenker, Stanford and Yang (PSSY) \cite{Penington:2019kki}. We will use this model below.  Consider an evaporating black hole in Jackiw-Teitelboim (JT) gravity with an end of the world (EOW) brane deep inside the black hole capping off the geometry, plus some propagating bulk degrees of freedom which constitute the code subspace. The Euclidean action is given by:
\begin{equation}
    I=I_{\mathrm{JT}}+\mu \int_{\text {brane }} \mathrm{d} s + I_{\text{field theory}},
\end{equation}
where $I_{\text{field theory}}$ is a bulk effective field theory action and the JT gravity action is 
\begin{equation}
    I_{\mathrm{JT}}=-\frac{S_{0}}{2 \pi}\left[\frac{1}{2} \int_{\mathcal{M}} \sqrt{g} R+\int_{\partial \mathcal{M}} \sqrt{h} K\right]-\left[\frac{1}{2} \int_{\mathcal{M}} \sqrt{g} \phi(R+2)+\int_{\partial \mathcal{M}} \sqrt{h} \phi (K-1)\right],
\label{eq:jt-action}
\end{equation}
with the boundary conditions:
\begin{equation}
    \left.\mathrm{d} s^{2}\right|_{\partial \mathcal{M}}=\frac{1}{\epsilon^{2}} \mathrm{~d} \tau^{2}, \quad \left.\phi\right|_{\partial\mathcal{M}}=\frac{1}{\epsilon}, \quad \epsilon \rightarrow 0.
\end{equation}
The boundary condition on the the EOW brane is given by
\begin{equation}
    \partial_{n} \phi=\mu\geq 0, \quad K=0.
\end{equation}
Here $S_0$ can be thought of as the entropy of the higher-dimensional extremal black hole for which the JT theory is the two-dimensional reduction, or it can be thought of as the ground state entropy of the JT system.

The EOW brane hosts semi-classical, intrinsic degrees of freedom, which we label by Greek letters, $\alpha,\beta$ etc.,  running from 1 to $k$. These intrinsic EOW brane states are taken to be orthogonal in the bulk semi-classical description i.e., these indices along the brane are contracted with the propagator $\delta_{\alpha\beta}$. 
We think of these states as arising from some sort of theory living on the brane, or alternatively as a flavor index for the brane which must be conserved along the worldline to ensure orthogonality.
The JT gravity black hole describes the semi-classical bulk dual to an excited state in a holographic quantum mechanical system, which we  label $B$. More precisely, JT gravity describes an ensemble average over the Hamiltonian of the dual quantum mechanical system $B$ \cite{Saad:2019lba}; in this context, the EOW brane states correspond to effectively random superpositions of energy eigenstates,\footnote{These superpositions also depend in a controlled way on the brane tension $\mu$ and inverse temperature set by Euclidean boundary length, which renders them not strictly random \cite{Penington:2019kki}.} with the coefficients chosen from a Gaussian ensemble with zero mean and unit variance \cite{Penington:2019kki}. 

In order to model an evaporating scenario, we entangle this system ($B$) with a quantum mechanical radiation system ($R$):
\begin{equation}
\label{eq: global states}
    \left|\Psi_i\right\rangle= \frac{1}{\sqrt{k}} \sum_{\alpha=1}^{k}\left|\psi^\alpha_i\right\rangle_B \otimes |\alpha \rangle_R,
\end{equation} 
where $\left|\psi^\alpha_i\right\rangle_B$ is a state in the quantum system $B$ whose semi-classical bulk dual consists of the black hole, including an EOW brane with its intrinsic degrees of freedom in the state $\alpha$, and the bulk code subspace fields in the code basis state $i$ ($i=1,...,d$).  Note that the entanglement is mediated by the EOW brane degrees of freedom, and not by the code subspace states. We will choose the code subspace to consist of simple, qubit-like excitations in the effective field theory, as will be explained below.   The strategy is to treat this one-parameter family (labelled by $k$) of entangled states as snapshots of an AdS black hole evaporating into a reservoir/bath, with $k$ playing the role of time.  More accurately, rather than considering an evaporating black hole {\it per se}, we maximally entangle a black hole with a fixed number of degrees of freedom with an ever larger number of radiation states and study the consequences. It was shown in \cite{Penington:2019kki} that as the parameter $k$ becomes large, there is a phase transition in the entanglement entropy $S(R)$ at $k \sim e^{S_0}$, where $S_0$ is the extremal black hole entropy. The phase transition cuts off the naive growth of entropy and realizes the expected Page curve. The mechanism by which this happens in the replica trick computation of  entropy involves the appearance of a new gravitational saddle, i.e., the replica wormhole, which makes the dominant contribution past the Page point \cite{Penington:2019kki}.

\section{Quantum error correction behind the horizon}\label{sec:qec-pssy}
We now come to the first main goal of this paper, which is to apply equation \eqref{SQEC} to the black hole interior in the PSSY toy model for an evaporating black hole. Following PSSY, we consider a black hole entangled with radiation \emph{past} the Page time:
\beq
|\Psi\rangle = \frac{1}{\sqrt{k}}\sum_{\alpha=1}^k |\psi^{\alpha}\rangle_B\otimes |\alpha\rangle_R,\;\;\;\cdots \;\;\;(k \gg e^{S_0})
\eeq
The physical Hilbert space is  $\cH_{\ph} = \cH_{B}\otimes \cH_R$, where $B$ is the microscopic quantum mechanical system describing the black hole, and $R$ is the Hilbert space of the radiation bath which absorbs the Hawking radiation. Loosely speaking, our code subspace will be the Hilbert space of bulk quantum field theory excitations on the black hole background, but we will specify it more precisely below. We will take this Hilbert space to factorize as
\beq
\mathcal{H}_{\text{code}} = \mathcal{H}_{\inte}\otimes \mathcal{H}_{\ext},
\label{eq:code-space-grav}
\eeq
i.e., it is a tensor product of the degrees of freedom in the black hole interior times degrees of freedom in the black hole exterior.
Let $\{|i\rangle_\inte \}$ be an orthonormal basis spanning the interior Hilbert space $\cH_{\inte}$ and let $\{|i'\rangle_\ext \}$ be an orthonormal basis spanning the exterior Hilbert space $\cH_{\ext}$. 
As before, there is a basis for $\cH_\co$ given by $\ket{i,i'}_\co = \ket{i}_\inte \otimes \ket{i'}_\ext$.
We view this system as a quantum error correcting code:
\beq
V: \mathcal{H}_{\inte}\otimes \mathcal{H}_{\ext} \to \mathcal{H}_{\ph}.
\eeq
Corresponding to a basis state $|i,i'\rangle_\co \in \cH_{\co}$, we have an image under the encoding in the physical Hilbert space:
\beq
|\Psi_{i,i'}\rangle_{\ph} =  V |i,i'\rangle_{\co} 
\eeq
given by
\beq
|\Psi_{i,i'}\rangle_\ph =\frac{1}{\sqrt{k}}\sum_{\alpha=1}^k |\psi^{\alpha}_{i,i'}\rangle_B\otimes |\alpha\rangle_R,
\eeq
where $|\psi^{\alpha}_{i,i'}\rangle_B$ is a state in the Hilbert space of the CFT dual to the black hole whose corresponding bulk geometry has an EOW brane in the internal state $\alpha$ and  bulk quantum fields in the state $|i,i'\rangle_{\co}$.  Thus, the encoding map leads to a representation of the code state as an excitation around a particular evaporating black hole (entangled with the radiation bath), which has a universal coarse description as a gravitating geometry, and a dual microscopic CFT representation. Past the Page time, the quantum extremal surface for the radiation bath lies at the bifurcation point in the black hole geometry, so that by the general entanglement wedge reconstruction paradigm, we expect the factor of the code subspace associated to the interior of the black hole to be encoded within the radiation Hilbert space.

\begin{figure}
    \centering
    \includegraphics[height=3.5cm]{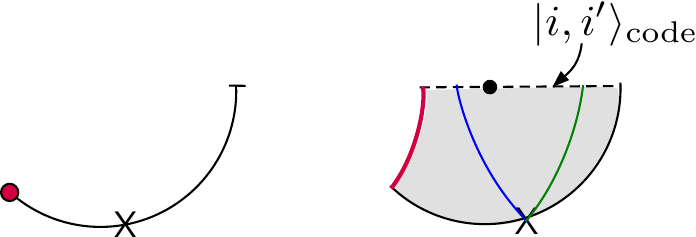}
    \caption{\small{\textbf{(Left)} The state $|\psi^{\alpha}_{i,i'}\rangle_B$ can be constructed in the CFT from a Euclidean path integral with appropriate superpositions of source configurations (indicated by a cross) and an EOW brane boundary condition $\alpha$ at $\tau = - \beta/2$ (red dot). \textbf{(Right)} The bulk description has a black hole with an EOW brane (shown in red) and bulk matter fields in the state $|i,i'\rangle_{\co}$. The black dot is the black hole bifurcation point.}}
    \label{fig:state}
\end{figure}

Formally, the state $|\psi^{\alpha}_{i,i'}\rangle_B$ can be constructed in the boundary CFT via a Euclidean path integral with sources turned on, corresponding to asymptotic boundary conditions for the bulk quantum fields (Fig.~\ref{fig:state}).  In Sec.~\ref{sec:non-iso}, we will show that this holographic encoding map is only an isometry up to non-perturbatively small $O(d e^{-S_0})$ corrections.  Therefore,  we can use the error-correcting code theorems discussed in Sec.~\ref{sec:QECreview} (which strictly speaking assume isometric encoding), so long as we are considering a sufficiently small code subspace with $d \ll e^{S_0}$. 

Another subtlety involves our assumption of a factorized code subspace in (\ref{eq:code-space-grav}). In the path integral construction of the encoding map, smooth sources producing finite energy densities will not result in states that are products between $\cH_{\inte}$ and $\cH_{\ext}$. Conversely, a state for bulk quantum fields that is a  product between the interior and exterior will generically have large energy density near the horizon.  However, we are interested in an effective field theory around the black hole background, which therefore comes equipped with a UV cutoff (much smaller than Planck scale). The Hilbert space thus does include product states between the interior and exterior, provided the excitations  are separated by distances much larger than the UV cutoff scale. For instance, we could imagine constructing a qubit degree of freedom in the interior and another qubit degree of freedom in the exterior, from sufficiently localized, low-energy operators in the effective field theory, well-separated from the horizon. Our analysis is performed in this setting.  By turning on various Euclidean sources on the asymptotic boundary \cite{Marolf:2017kvq}, we can at least formally construct such states $|\psi^{\alpha}_{i,i'}\rangle_B$ corresponding to low energy bulk modes of the effective field theory in product states $|i,i'\rangle_{\co}$.  We will ignore $O(G_N)$ back-reaction corrections to the geometry and the state of bulk matter fields in the code subspace.

As mentioned before, past the Page time, the black hole interior is in the entanglement wedge of the radiation/bath. We will ask what type of quantum operations acting on the bath Hilbert space $\mathcal{H}_R$ are correctable on the code subspace in the black hole interior. By correctable, we mean that we should be able to recover the reduced density matrix of bulk modes in the interior of the black hole (i.e., with support on $\cH_{\inte}$) after the action of such errors. 

\subsection{Coherent information criterion}
The question we wish to address fits perfectly within the formalism of subsystem quantum error correction explained in the previous section. Following that logic, we model the error as a quantum channel $\cE=\{E_m\}$ which acts on the physical Hilbert space $\cH_{B}\otimes \cH_{R}$. More precisely, we will only consider errors which act on the radiation bath Hilbert space $\cH_R$, leaving $\cH_B$ untouched. Next, we introduce two reference systems $\fr_i$ and $\fr_e$ which have the same dimensions as $\mathcal{H}_{\inte}$ and $\mathcal{H}_{\ext}$ respectively.  
Recall that $\cH_\inte$ and $\cH_\ext$ are the two factors of $\cH_\co$. 
Now define the state
\beq
|\Psi'\rangle = \sum_{i,i',m} \ket{i,i'}_\re \otimes  E_m|\Psi_{i,i'}\rangle_{\ph} \otimes \ket{e_m}_\en
\eeq
where, following the discussion in Sec.~\ref{sec:QECreview}, we have introduced another factor $\mathcal{H}_\en$ corresponding to an auxiliary environment, and the reference space is $\mathcal{H}_\re = \mathcal{H}_{\fr_i} \otimes \mathcal{H}_{\fr_e}$. In order to determine whether the reduced density matrix of bulk quantum fields in the black hole interior is correctable under the action of $\cE$, we are instructed to compute the mutual information:
\beq
I_{\Psi'}(\fr_i: \fr_e\cup \en) \equiv S_{\Psi'}(\fr_i) + S_{\Psi'}(\fr_e\cup \en) - S_{\Psi'}(\fr_i \cup \fr_e \cup\en).
\label{eq:mutualinfo}
\eeq
The error is correctable if and only if $I_{\Psi'}(\fr_i: \fr_e\cup \en) = 0$, i.e., the corresponding density matrix factorizes as 
\beq
\rho'_{\fr_i,\fr_{e},\en} = \rho'_{\fr_i} \otimes \rho'_{\fr_{e},\en}.
\eeq
For gravity computations, it is often more convenient to consider the R\'enyi version of the above mutual information:
\beq
I^{(n)}_{\Psi'}(\fr_i: \fr_e\cup \en) \equiv S^{(n)}_{\Psi'}(\fr_i) + S^{(n)}_{\Psi'}(\fr_e\cup \en) - S^{(n)}_{\Psi'}(\fr_i \cup \fr_e \cup\en).
\label{eq:renyimutualinformation}
\eeq
More precisely, instead of computing the von Neumann entropies $S_{vN}=-\mathrm{Tr}\,\rho \log \rho$, we compute the R\'enyi entropy $S^{(n)} = \frac{1}{1-n}\log\,\mathrm{Tr}\,\rho^n$ for integer values of $n$. The von Neumann entropies can then be obtained by analytically continuing the R\'enyi entropies in $n$ and taking the limit $n\to 1^+$. So, our goal now is to compute the above R\'enyi mutual information. 

\begin{figure}[t]
    \centering
    \includegraphics[height=5cm]{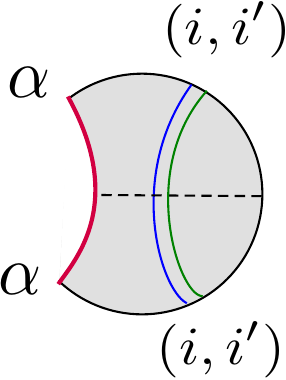}
    \caption{\small{The Euclidean black hole geometry which enters in determining the normalization factor $\cN$ defined in \eqref{eq:normalization-defn}, specifically contributing to the overlap $\inner{\psi_{i,i'}^\alpha}{\psi_{i,i'}^\alpha}_B$.  The red line is the EOW brane with index $\alpha$, the blue line is an interior code subspace excitation with label $i$, and the green line is an exterior code subspace excitation with label $i'$. }}
    \label{fig:BH}
\end{figure}

In order to compute the R\'enyi mutual information, we first compute the reduced density matrix on $\fr_i \cup \fr_e \cup \en$ by tracing out the physical Hilbert space:
\beq\label{redDM}
\rho'_{\fr_i,\fr_e,\en} = \frac{1}{k\mathcal{N}} \sum_{i,i',j,j',m,n}|i,i',e_m\rangle\langle j,j',e_n|_{\fr_i,\fr_e,\en}\sum_{\alpha,\beta}\langle \psi^{\alpha}_{j,j'}|\psi^{\beta}_{i,i'}\rangle_B \langle \alpha| E_n^{\dagger}E_m | \beta\rangle_{R},
\eeq
where $\mathcal{N}$ is a normalization constant.   To determine the normalization constant, we take the trace of the above expression. Using equation \eqref{trC}, we get
\beq
\mathrm{Tr}\,\rho'_{\fr_i,\fr_e,\en} = \frac{1}{k\mathcal{N}} \sum_{i,i'}\sum_{\alpha}\langle \psi^{\alpha}_{i,i'}|\psi^{\alpha}_{i,i'}\rangle_B 
\label{eq:normalization-defn}
\eeq
where  $i$ and $i'$ are indices in the code space and $\alpha$ is an index of the EOW brane state.
Using the gravity description in the semi-classical limit, the overlap is computed by the geometry depicted in Fig.~\ref{fig:BH}.
This geometry is filling in a boundary condition which consists of two copies of Fig.~\ref{fig:state} that are glued together to produce the overlap
\beq
\langle \psi^{\alpha}_{i,i'}|\psi^{\alpha}_{i,i'}\rangle_B= e^{S_0}Z_1 \langle i,i'|i,i'\rangle_{\co}= e^{S_0}Z_1 ,
\eeq
where $Z_1$ is the exponential of the on-shell gravitational action (\ref{eq:jt-action}) (or beyond the semi-classical approximation, the JT gravity path integral) on a disc capped off by an EOW brane, along with the non-code-subspace bulk field theory modes in their Hartle-Hawking vacuum state. We explicitly pulled out the factor of $e^{S_0}$ coming from the first term in the JT action.  Since we are ignoring the back-reaction from the effective field theory, so on the disc it simply gives the path integral overlap $\langle i, i' | i,i'\rangle_\co$.
Then, enforcing $\Tr \rho'_{\fr_i,\fr_e,\en} = 1$ determines the normalization factor
\beq
\mathcal{N} = d_{i}d_ee^{S_0}Z_1,
\label{eq:norm-density-matrix}
\eeq
where $d_i$ and $d_e$ are the dimensions of the interior and exterior factors in the code subspace respectively.
We will continue to make these contributions of the code subspace portion of the effective field theory explicit, and absorb the remainder of the bulk field theory modes into partition functions like $Z_1$.

Let us now evaluate $S^{(n)}_{\Psi'}(\fr_i \cup \fr_e \cup \en)$. From equation \eqref{redDM}, we first obtain
\beqn
\mathrm{Tr}\,{\rho'}^n_{\fr_i,\fr_e,\en} &=& \frac{1}{k^n\mathcal{N}^n} \sum_{i_1,\cdots,i_n}\sum_{i_1',\cdots,i'_n}\sum_{\alpha_1\cdots,\alpha_n}\sum_{\beta_1,\cdots,\beta_n}\sum_{m_1,\cdots, m_n}\langle \psi^{\alpha_1}_{i_2,i_2'}|\psi^{\beta_1}_{i_1,i_1'}\rangle_B \cdots \langle \psi^{\alpha_n}_{i_1,i_1'}|\psi^{\beta_n}_{i_{n},i_{n}'}\rangle_B\nonumber\\
&\times & \langle \alpha_1| E_{m_2}^{\dagger}E_{m_1}|\beta_1\rangle_R\cdots \langle\alpha_n| E_{m_1}^{\dagger}E_{m_n}|\beta_n\rangle_R.
\eeqn
Here $i$ and $i'$ are indices in the code subspace, $\alpha$ and $\beta$ are EOW brane indices, and the $m$ indices label Kraus operators.  
Notice that the reference space and auxiliary environment have disappeared here because we are taking the trace on them.  It is convenient to group together all the factors pertaining to the channel $\cE$ and the corresponding sums over the environment indices to define a tensor which we will call $f$:
\beq
f_{\alpha_1,\beta_1,\cdots \alpha_n,\beta_n}  \equiv  \sum_{m_1,\cdots, m_n}\langle \alpha_1| E_{m_2}^{\dagger}E_{m_1}|\beta_1\rangle_R\cdots \langle\alpha_n| E_{m_1}^{\dagger}E_{m_n}|\beta_n\rangle_R.
\eeq
We can re-express the $f$-tensor in terms of the isometric extension of $\cE$ as
\begin{equation}
\begin{split}
f_{\alpha_1,\beta_1,\cdots \alpha_n,\beta_n}  & = \sum_{m_1,\cdots, m_n}\sum_{\gamma_1,\cdots, \gamma_n}\langle \alpha_1,e_0|U_\cE^{\dagger}| \gamma_1,e_{m_2}\rangle_{R,\en}  \langle \gamma_1,e_{m_1}|U_\cE |\beta_1,e_0\rangle_{R,\en} \cdots \\
& \quad \times \langle \alpha_n,e_0|U_\cE^{\dagger}| \gamma_n, e_{m_1}\rangle_{R,\en}   \langle \gamma_n, e_{m_n}|U_\cE |\beta_n,e_0\rangle_{R,\en}  ,
\end{split}
\label{eq:f-tensor}
\end{equation}
where in the second sum we have also inserted complete sets of radiation states $\sum_\gamma \ket{\gamma}\bra{\gamma}_R = \mathbb{1}_R$.
In terms of the $f$-tensor defined above, we have
\beq \label{tr1}
\mathrm{Tr}\,{\rho'}^n_{\fr_i,\fr_e,\en} = \frac{1}{k^n\mathcal{N}^n} \sum_{i_1,\cdots,i_n}\sum_{i_1',\cdots,i'_n}\sum_{\alpha_1\cdots,\alpha_n}\sum_{\beta_1,\cdots,\beta_n}\langle \psi^{\alpha_1}_{i_2,i_2'}|\psi^{\beta_1}_{i_1,i_1'}\rangle_B \cdots \langle \psi^{\alpha_n}_{i_1,i_1'}|\psi^{\beta_n}_{i_{n},i_{n}'}\rangle_B\,f_{\alpha_1,\beta_1,\cdots,\alpha_n,\beta_n}.
\eeq

\begin{figure}
    \centering
    \begin{tabular}{ c c}
    \includegraphics[height=6cm]{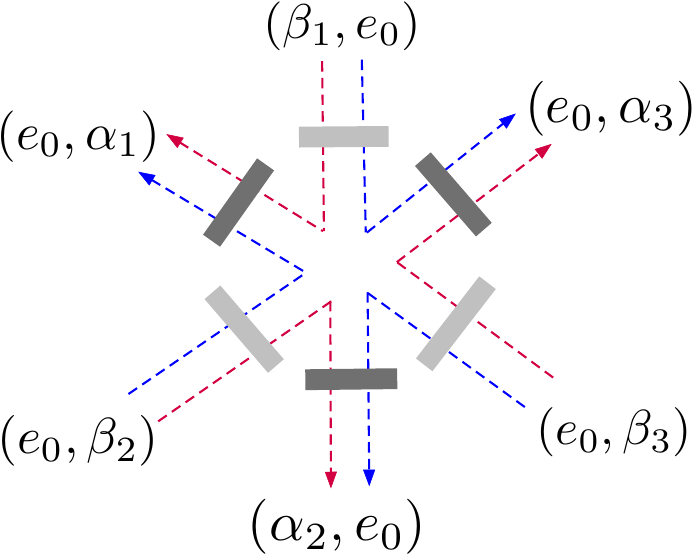} & \includegraphics[height=6cm]{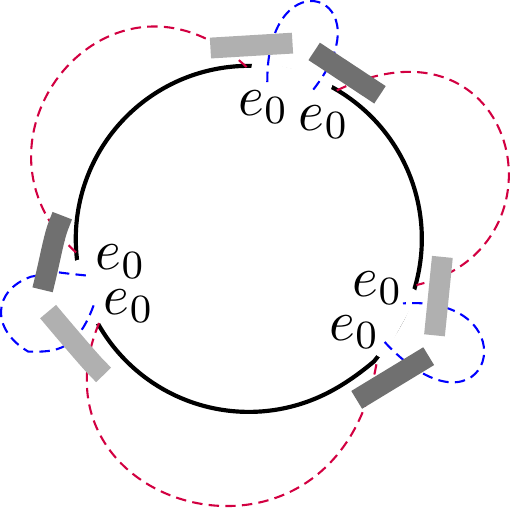}
    \end{tabular}
    \caption{\small{\textbf{(Left)} A tensor network representation of the index structure $f_{\alpha_1,\beta_1,\cdots,\alpha_n,\beta_n}$ with $n=3$. The light grey boxes denote $U_\cE$ while dark grey boxes denote $U_\cE^{\dagger}$.  Dashed blue and red internal legs denote environment and radiation indices respectively.  External legs are outgoing (incoming) to denote uncontracted indices in a bra (ket). \textbf{(Right)} The contraction of the $f$-index structure with the gravitational overlaps (i.e., asymptotic boundaries), denoted with solid black lines, required to compute the sums in (\ref{tr1}, \ref{tr2}).  Each term in these sums involves a product of overlaps between $n$ sets of asymptotic gravitational states.  The states are defined in terms of operators acting at the asymptotic boundaries which are illustrated here as black lines.  Below we will see that the gravitational path integral can fill in these boundaries in various ways that may connect them or not through a bulk geometry. 
    }
    }
    \label{fig:f}
\end{figure}

Using the isometric extension, the $f$-tensor can be conveniently visualized in terms of a tensor network, as shown in Fig.~\ref{fig:f}. Similarly, we get
\beq \label{tr2}
\mathrm{Tr}\,{\rho'}^n_{\fr_e,\en} = \frac{1}{k^n\mathcal{N}^n} \sum_{i_1,\cdots,i_n}\sum_{i_1',\cdots,i'_n}\sum_{\alpha_1\cdots,\alpha_n}\sum_{\beta_1,\cdots,\beta_n}\langle \psi^{\alpha_1}_{i_1,i_2'}|\psi^{\beta_1}_{i_1,i_1'}\rangle_B \cdots \langle \psi^{\alpha_n}_{i_n,i_1'}|\psi^{\beta_n}_{i_{n},i_{n}'}\rangle_B\,f_{\alpha_1,\beta_1,\cdots,\alpha_n,\beta_n}.
\eeq
Finally,
\beq\label{tr3}
\mathrm{Tr}\,{\rho'}^n_{\fr_i} = \frac{1}{k^n\mathcal{N}^n} \sum_{i_1,\cdots, i_n}\sum_{i_1',\cdots,i_n'}\sum_{\alpha_1,\cdots, \alpha_n}\langle \psi^{\alpha_1}_{i_2,i_1'}|\psi^{\alpha_1}_{i_1,i_1'}\rangle_B\cdots \langle\psi^{\alpha_n}_{i_1,i_n'}|\psi^{\alpha_n}_{i_n,i_n'}\rangle_B.
\eeq
The coefficients in these various expressions involving products of overlaps of the black hole states can be computed using the standard rules of JT gravity. 

More precisely, JT gravity plus EOW branes computes an \emph{ensemble average} of microscopic theories \cite{Saad:2019lba,Penington:2019kki,Gao:2021uro}, but the error correction theorems we developed in Sec.~\ref{sec:QECreview}, refer to single theories, not an ensemble.  We can nevertheless use the JT gravity approach because, in the limit of large $k$ and $S_0$, the ensemble average is a good approximation to the typical answer since the variance is suppressed in $k$ and $e^{-S_0}$. We must carefully to account for the various different bulk geometries which can fill in the fixed asymptotic boundary conditions. Below, we will first consider the two most important geometries, namely the fully disconnected geometry  and the fully connected one; both of these preserve the $\mathbb{Z}_n$ symmetry of the asymptotic boundary conditions.  We will also argue later that at least for erasure errors, the $\mathbb{Z}_n$-breaking bulk geometries do not dominate, at least far from the phase transition between the two $\mathbb{Z}_n$-symmetric solutions.  However, we do not have a general proof for all errors. 

\subsection*{An assumption about $\cE$}

In the computations that follow, we will assume the error operation $\cE$ does not have prior access to black hole microstates, or more precisely, the $f$-tensor structure, which contains all the details of the quantum channel $\cE$, has no additional black hole microstate overlaps in it. Thus, we will only have to evaluate the products of gravitational overlaps $\langle \psi^{\alpha}_{i,i'}|\psi^{\beta}_{j,j'}\rangle_B$ that are explicitly written in equations \eqref{tr1}, \eqref{tr2} and \eqref{tr3} using JT gravity.   If the assumption were not true, then the evaluation of the $f$-tensor would contain additional gravitational path integrals, and we would have to account for additional geometries in JT gravity which potentially connect between the overlaps appearing explicitly in (\ref{tr1},\ref{tr2},\ref{tr3})  and those contained inside the $f$-tensor structure.
A simple example of this situation would be if the Stinespring dilation of the channel $\cE$ couples an external environment qubit to the Petz operator reconstruction on the bath of a bulk operator in the black hole interior \cite{Penington:2019kki}.  Since the Petz operator is constructed with knowledge of black hole microstates (or more precisely, with knowledge of the encoding map $V$), this channel would violate our assumption.\footnote{Such channels may be related to the ``miracle operators" discussed in \cite{Qi:2021sxb}.}

Furthermore, we expect such channels to also be exponentially complex in the sense of circuit complexity.  Indeed, \cite{Kim:2020cds} conjectured that the density matrix on the bath Hilbert space is \emph{pseudorandom}, i.e., it cannot be reliably distinguished from the maximally mixed state on the bath with polynomial complexity operations. 
It would be interesting to better understand the interplay between this pseudorandom assumption and our assumption about the absence of microstate overlaps in $\cE$.

\subsection*{Fully disconnected geometry}

\begin{figure}
    \centering
    \includegraphics[height=4.5cm]{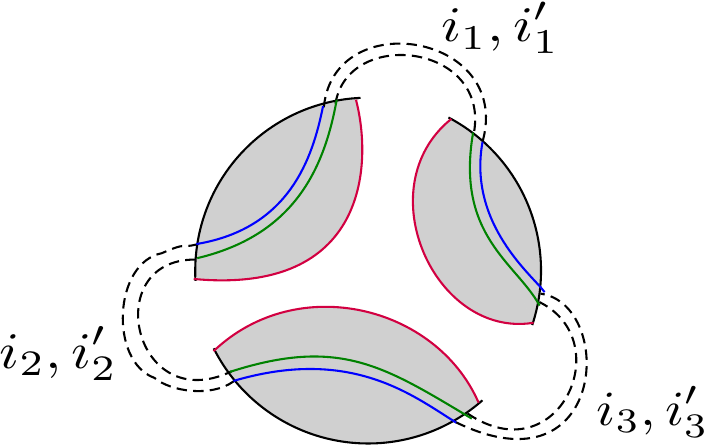} 
    \caption{\small{The disconnected gravitational contribution to the R\'enyi entropy of $\fr_i\cup \fr_e \cup \en$. The grey region is the gravitational geometry filling in the boundary conditions, while the red lines are EOW branes.  Green and blue lines denote bulk contractions of the exterior and interior factors in the code subspace respectively. Dotted lines indicate identifying and summing over the code-subspace indices.}}
    \label{fig:discode1}
\end{figure}

\begin{figure}
    \centering
    \begin{tabular}{c c}
    \includegraphics[height=5.5cm]{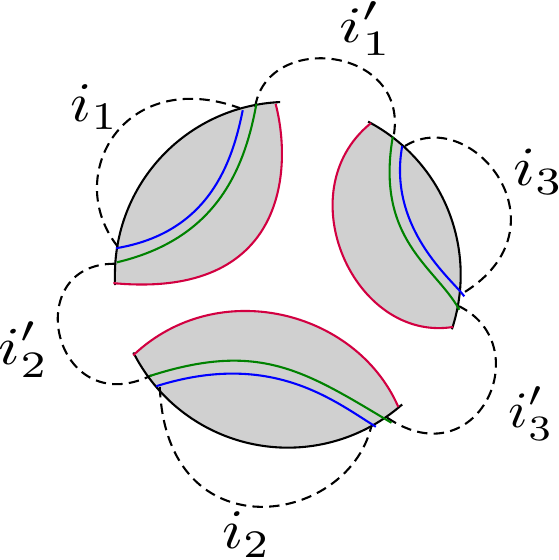} & \includegraphics[height=5.5cm]{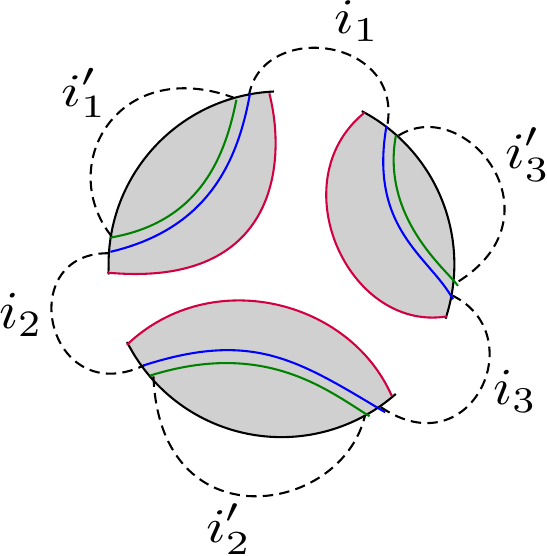}
    \end{tabular}
    \caption{\small{The disconnected gravitational contributions to the R\'enyi entropies of $\fr_e \cup\en$ \textbf{(left)} and $\fr_i$ \textbf{(right)}. The grey region is the gravitational geometry filling in the boundary conditions, while the red lines are EOW branes.  Green and blue lines denote bulk contractions of the exterior and interior factors in the code subspace respectively. }}
    \label{fig:discode2}
\end{figure}

We now turn to the evaluation of the various entropies in JT gravity. To begin we consider the fully disconnected contribution where each of the gravitational overlaps in (\ref{tr1}), (\ref{tr2}) and (\ref{tr3}) is evaluated by a separate bulk geometry. In this case, we can simply use the formula
\beq
\langle \psi^{\alpha}_{i,i'}|\psi^{\beta}_{j,j'}\rangle_B= e^{S_0}Z_1 \delta_{\alpha\beta}  \delta_{ij}\delta_{i'j'} .
\label{eq:DiscOverlap}
\eeq
This is derived in an identical fashion to the calculation of the norm $\cN$ in Fig.~\ref{fig:BH}, but with the more general code subspace overlap $\inner{i,i'}{j,j'}_\co = \delta_{ij}\delta_{i'j'}$ and EOW brane index matching yielding $\delta_{\alpha\beta}$.  The resulting computation is diagrammed in Figs.~\ref{fig:discode1} and \ref{fig:discode2}, and gives
\beq
\mathrm{Tr}\,{\rho'}^n_{\fr_i,\fr_e,\en}\Big|_{\text{disconn.}} =  d^{1-n}_id^{1-n}_e\;\mathrm{Tr}\,\sigma^n_{\en},
\label{eq:disc-result1}
\eeq
\beq
\mathrm{Tr}\,{\rho'}^n_{\fr_e,\en}\Big|_{\text{disconn.}} =  d^{1-n}_e\;\mathrm{Tr}\,\sigma^n_{\en},
\label{eq:disc-result2}
\eeq
\beq
\mathrm{Tr}\,{\rho'}^n_{\fr_i}\Big|_{\text{disconn.}} =  d^{1-n}_i,
\label{eq:disc-result3}
\eeq
where we have defined
\beq
\sigma_{\en} = \mathrm{Tr}_{R} \left\{ U_\cE\left(\frac{1}{k}\sum_{\alpha=1}^k |\alpha\rangle\langle\alpha|_R \otimes |e_0\rangle\langle e_0|_{\en}\right) U_\cE^{\dagger}\right\}.
\label{eq:sigmaenv}
\eeq
For example, the code subspace dimension factors in the disconnected contribution to $\Tr {\rho'}^n_{\fr_i,\fr_e,\en}$ are computed by observing in Fig.~\ref{fig:discode1} that there is one green index loop yielding $d_e$, one blue index loop yielding $d_i$, and the normalization $\cN$ contributes $(d_id_e)^{-n}$.
The $\mathrm{Tr}\,\sigma^n_{\en}$ contribution comes from the $f$-tensor with appropriate index contractions dictated by the EOW brane propagators (Kronecker deltas matching indices) in the geometry. 

The contribution to the R\'enyi mutual information coming from the fully disconnected geometry is given by taking the logarithm of \eqref{eq:disc-result1} and subtracting from it the logarithm of \eqref{eq:disc-result2} plus the logarithm of \eqref{eq:disc-result3}:
\beq
I^{(n)}_{\Psi'}(\fr_i: \fr_e, \en)\Big|_{\text{disconn.}}=0.
\label{eq:disconnMI}
\eeq
Thus, when the disconnected geometry dominates the gravity path integral,  the encoding of the black hole interior is robust, i.e., error channels that act only on the bath $\cH_R$ do not affect the encoding of the black hole interior. Of course, the mutual information is not strictly zero, because of sub-leading in $e^{-S_0}$ contributions.
Nevertheless, the state on the black hole interior is at least approximately correctable, i.e., there exists a recovery channel $\cR$ such that
\beq
\text{max}_{\rho}|| \cR( \cE(\rho)) - \rho ||_1 \leq \epsilon,\;\;\;\epsilon \sim e^{-S_0/4},
\eeq
where the maximization is over all density matrices supported on the code subspace.
This result follows from the error correction theorems we reviewed in Sec.~\ref{sec:QECreview}, specifically \eqref{eq:mutual-bound}.

\subsection*{Fully connected geometry}
\begin{figure}[t]
    \centering
    \begin{tabular}{c c c}
         \includegraphics[height=4.8cm]{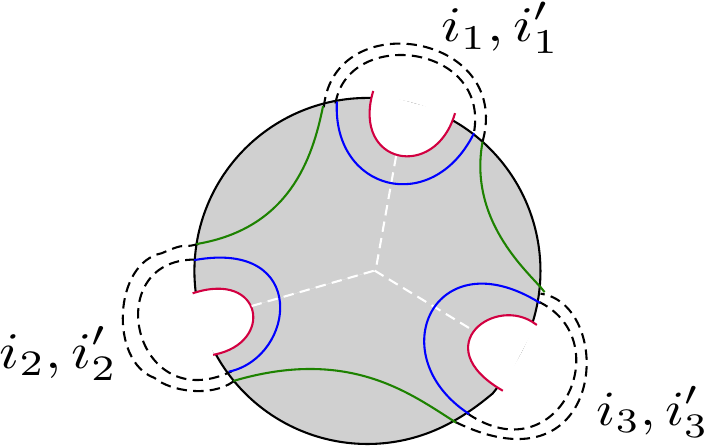} & \hspace{0.2cm} &  \includegraphics[height=4.4cm]{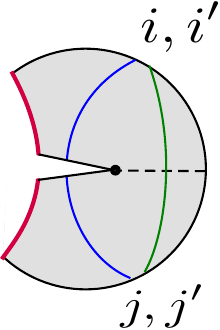} 
    \end{tabular}
   \caption{\small{\textbf{(Left)} The connected gravitational contribution to the R\'enyi entropy of $\fr_i\cup \fr_e \cup \en$. Green and blue lines denote bulk contractions of the exterior and interior factors in the code subspace respectively.  The dashed white lines separate fundamental  domains of the $\mathbb{Z}_3$ replica symmetry. \textbf{(Right)} The $\mathbb{Z}_n$ quotient of the replica wormhole in the $n\to 1^+$ limit.  The geometry of this manifold, with a cut on the interior, provides a specific prescription for the contractions of the bulk code subspace indices.
   The bulk field theory state created on the $\mathbb{Z}_2$-symmetric slice is replicated and results in a bulk R\'enyi entropy for the black hole interior.
   In order to compute this entropy, the interior code subspace excitation (blue line) must pass between replicas and the exterior code subspace excitation (green line) must remain within a given replica.
   This pattern of contractions is drawn on the left figure.
   The images of the cut on the $n\to 1^+$ geometry are the white dashed lines on the $n=3$ geometry.
   }}
    \label{fig:conncode1}
\end{figure}

\begin{figure}
    \centering
    \includegraphics[height=5.6cm]{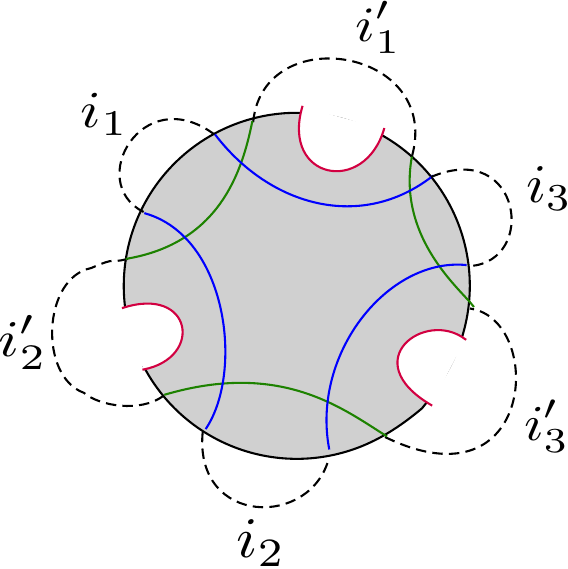} 
    \caption{\small{The connected gravitational contribution to the R\'enyi entropy of $\fr_e \cup \en$. The grey region is the  manifold which fills in the boundary conditions defined by the gravitational overlaps in \eqref{tr2}.  The red lines are the EOW branes. Green and blue lines denote bulk contractions of the exterior and interior factors in the code subspace respectively. }}
    \label{fig:conncode2}
\end{figure}

Let us now consider the fully connected contribution. This geometry is the analog of the replica wormhole saddle. By explicit computation, one can prove that this contribution never dominates in $\mathrm{Tr}\,{\rho'}^n_{\fr_i}$, and the dominant contribution is always the disconnected geometry. However, in the other two cases -- i.e., $\mathrm{Tr}\,{\rho'}^n_{\fr_i,\fr_e,\en}$ and $\mathrm{Tr}\,{\rho'}^n_{\fr_e,\en}$ -- the replica wormhole (Figs.~\ref{fig:conncode1} and \ref{fig:conncode2}) can become dominant, as we will see below by explicit computation.  In these cases, the gravitational path integral for general $n$ result is not easy to compute.  But because we are interested in taking the $n\to 1^+$ limit we can use the procedure proposed by Lewkowycz and Maldacena \cite{Lewkowycz:2013nqa}: (1) assume that the relevant geometries that contribute are the replica symmetric ones, (2) 
compute the gravitational action of the  $\mathbb{Z}_n$ quotient of these geometries (Figs.~\ref{fig:conncode1} and~\ref{fig:conncode2}, as shown on the right in Fig.~\ref{fig:conncode1}), and (3) analytically continue to $n=1^+$.

Applying this procedure in the $n\to 1^+$ limit, the quotient geometry is (approximately) identical to a copy of a single Euclidean black hole, but with a conical singularity at the black hole bifurcation point, and in addition a twist operator insertion for the bulk fields which corresponds to a specific prescription for contracting the bulk code subspace indices, as shown in Figs.~\ref{fig:conncode1} and \ref{fig:conncode2}. 
For example, the gravitational contribution for the connected phase of $\Tr {\rho'}^n_{\fr_i,\fr_e,\en}$
 is computed by evaluating Fig.~\ref{fig:conncode1}, and this gives 
\begin{equation}
    \frac{1}{k^n\cN^n} \sum_{i_1,\dots , i_n} \sum_{i'_1,\dots , i'_n} \inner{\psi^{\alpha_1}_{i_2,i_2'}}{\psi^{\beta_1}_{i_1,i_1'}} \dots \inner{\psi^{\alpha_n}_{i_1,i_1'}}{\psi^{\beta_n}_{i_n,i_n'}} = e^{S_0(1-n)} \frac{Z_n}{k^nZ_1^n}  d_e^{1-n} \delta_{\alpha_1\beta_2} \dots \delta_{\alpha_n\beta_1} ,
\label{eq:conn-EOW-structure}
\end{equation}
where $Z_n$ is the exponential of the on-shell gravity action on the disk geometry with $n$ EOW branes, times the bulk field theory path integral on this geometry.
There is a similar formula for the gravitational contribution to the connected phase of $\Tr {\rho'}^n_{\fr_e,\en}$, with the only difference being that we get a factor of $d_i^{1-n}d_e^{1-n}$
instead of $d_e^{1-n}$ because of the different structure of the bulk code subspace excitation loops (blue and green loops in Figs.~\ref{fig:conncode1} and \ref{fig:conncode2}).

Putting everything together, we get 
\beq
\mathrm{Tr}\,{\rho'}^n_{\fr_i,\fr_e,\en}\Big|_{\text{conn.}} \sim  e^{(1-n)[S_0+\frac{2\pi}{\beta}\phi_0+s_{\text{bulk}}]}d^{1-n}_e\;\mathrm{Tr}\,\sigma^n_{R},
\label{eq:re-env-conn1}
\eeq
\beq
\mathrm{Tr}\,{\rho'}^n_{\fr_e,\en}\Big|_{\text{conn}}\sim e^{(1-n)[S_0+\frac{2\pi}{\beta}\phi_0+s_{\text{bulk}}]}d^{1-n}_id^{1-n}_e\;\mathrm{Tr}\,\sigma^n_{R},
\label{eq:re-env-conn}
\eeq
where $\phi_0$ is the value of the dilaton at the bifurcation point, and $s_{\text{bulk}}$ is the bulk entropy for all the bulk modes in their Hartle-Hawking vacuum (i.e., those which are not excited in the code subspace).   In the above formulas, we have used $Z_n/(Z_1)^n\sim e^{(1-n)[S_0 + (2\pi/\beta) \phi_0 + s_{\text{bulk}}]}$ because we are working in the approximation that $(n-1)$ is small and positive, where the left hand side can be evaluated semi-classically using the Lewkowycz-Maldecena method. 
The factors of $S_0$ and $\phi_0$ come from JT gravity, while the $s_{\text{bulk}}$ comes from the bulk field theory.  
The density matrix $\sigma_R$  is defined as 
\beq
\sigma_{R} = \mathrm{Tr}_{\en} \left\{ U_\cE\left(\frac{1}{k}\sum_{\alpha=1}^k |\alpha\rangle\langle\alpha|_R \otimes |e_0\rangle\langle e_0|_{\en}\right) U_\cE^{\dagger}\right\} = \cE \left( \frac{1}{k}\mathbb{1}_R \right).
\label{eq:sigmaR}
\eeq
The second equation follows because, upon tracing out the environment factor from the action of the Stinespring dilation, we obtain the action of the channel $\cE$.
The particular state upon which the channel acts is the maximally mixed radiation state, i.e., the ``semi-classical state'' on the radiation.
As before, this $\mathrm{Tr}\,\sigma^n_{R}$ contribution comes from the $f$-tensor with appropriate index contractions dictated by the EOW brane propagators in the connected geometry.
Notice that here the reduced state $\sigma_R$ has appeared, whereas in the disconnected geometry it was $\sigma_\en$.
This difference has to do with the manner in which the EOW brane indices are contracted, and is demonstrated by explicit calculation.


We can take logarithms of the above formulae to show that when the fully connected geometry dominates in the  R\'enyi entropies we get a non-trivial mutual information 
\beq
I_{\Psi'}(\fr_i: \fr_e, \en)\Big|_{\text{conn.}}=2 \log d_i.
\label{eq:connMI}
\eeq
Therefore, in this case the bulk density matrix in the black hole interior is not correctable under the action of the error on the bath. The mechanism underlying this failure of error correction in gravity is the appearance of the replica wormhole. 

Finally, we are interested in the conditions for the fully connected geometry to dominate over the disconnected one.
For $\mathrm{Tr}\,{\rho'}^n_{\fr_e,\en}$, the criterion for the connected geometry to be significantly greater than the disconnected one when $(n-1)$ is small and positive is 
\beq
\frac{\mathrm{Tr}\,\sigma^n_{\en}}{\mathrm{Tr}\,\sigma^n_{R}} \ll  e^{(1-n)[S_{BH} + \log d_i]} .
\label{eq:entropy-diff}
\eeq
The analogous criterion for $\mathrm{Tr}\,{\rho'}^n_{\fr_i,\fr_e,\en}$ is
\begin{equation}
    \frac{\Tr \sigma_\en^n}{\Tr \sigma_R^n} \ll e^{(1-n)[S_{BH} - \log d_i]} ,
\label{eq:correctability}
\end{equation}
where we have defined $S_{BH} = S_0+\frac{2\pi}{\beta}\phi_0 + s_{\text{bulk}}$. (As before, $S_0$ is the extremal entropy, $\frac{2\pi}{\beta}\phi_0$ is the classical horizon contribution, and  $s_{\text{bulk}}$ is the entropy of bulk quantum fields.) 
Notice that the first inequality actually implies the second, so if $\mathrm{Tr}\,{\rho'}^n_{\fr_e,\en}$ is dominated by the connected geometry then so is $\mathrm{Tr}\,{\rho'}^n_{\fr_i,\fr_e,\en}$, but the converse is not true except when $d_i = 1$.
If we only have $\mathrm{Tr}\,{\rho'}^n_{\fr_i,\fr_e,\en}$ dominated by the connected geometry, the above equations show we still have $I_{\Psi'} \neq 0$.
Therefore, we ought to characterize correctability only by the connected phase of $\mathrm{Tr}\,{\rho'}^n_{\fr_i,\fr_e,\en}$.
Taking the logarithm on both sides of \eqref{eq:correctability}, multiplying by $\frac{1}{n-1}$, and taking $n\to 1^+$, we get
\beq\label{ci}
S(\sigma_{R})-S(\sigma_{\en}) \ll -( S_{BH}- \log d_i),
\eeq
The left hand side above is entirely a property  of the quantum channel $\cE$, with no reference to gravity. In quantum information theory, this quantity is referred to as the \emph{coherent information}.

For a quantum channel $\cE$ which can be isometrically extended to a unitary $U_\cE$ acting jointly on the system plus an environment factor $\cH_{\en}$, the coherent information of the channel $\cE$ with respect to an input state $\rho$ is defined as:
\beq
I_c(\cE;\rho) \equiv  S(\cE(\rho))-S(\rho'_{\en}),
\label{eq:coherentinfo}
\eeq
where $\rho'_{\en}$ is the reduced density matrix of the environment after the action of the joint unitary $U_\cE$. 
The coherent information is a measure for how much of the quantum information in the input state survives the action of the quantum channel; for instance, it decreases monotonically under composition of quantum channels. For our purposes, the input state is the maximally mixed state on $R$, which we can loosely think of as the ``semi-classical state'' of the radiation, i.e., a coarse-grained description of the state of the radiation obtained by keeping track of the coupled evolution of the bath with the black hole in the semi-classical gravity description.\footnote{In contrast, the microscopic state of the radiation obtained by keeping track of the coupled evolution with the quantum system dual to the black hole will not be maximally mixed/thermal and will have subtle correlations which restore unitarity.} Equation \eqref{ci} therefore states that if a quantum channel has a sufficiently negative coherent information with respect to the semi-classical state on the radiation Hilbert space, then the encoding of the black hole interior is no longer correctable under such a channel. Conversely, the encoding of the bulk state in the black hole interior is robust against the action of any quantum channel satisfying
\beq \label{CIcond}
I_c \left( \cE;\frac{1}{k}\mathbb{1} \right) \gg -( S_{BH}-\log d_i),
\eeq
together with our previous assumption that the channel should not alter the asymptotic Euclidean boundary conditions by sourcing additional boundaries. Therefore, as long as the remaining black hole is macroscopic, we find that the black hole interior is robustly encoded in the radiation. 
If the black hole entropy is instead so small as to be comparable to the $O(1)$ code subspace entropy $\log d_i$, an error with very small negative coherent information could corrupt the interior encoding. In either case, once the bound \eqref{CIcond} is violated, the black hole interior moves over to the entanglement wedge of the environment factor $\cH_{\en}$ and QEC on the bath fails at that point. 

It is worth emphasizing that a channel with large, negative coherent information need not be very complex. Consider the following example. Let the dimension of the environment factor be the same as the dimension of the bath Hilbert space, i.e., $\mathrm{dim}(\cH_{R}) = \mathrm{dim}(\cH_{\en})=k= 2^M$, and let the unitary $U$ correspond to the \text{SWAP} operator. For such a channel, we have
\beq
S(\sigma_R) = 0,\;\;\;S(\sigma_\en) = \log\,k = M \log 2,
\eeq
and so the coherent information is given by
\beq
I_c \left( \cE;\frac{1}{k}\mathbb{1} \right) = -M \log 2 \ll -(S_{BH}- \log\,d_i).
\eeq
Clearly, the SWAP channel violates the bound \eqref{CIcond}. This is not at all surprising because this channel essentially  transfers all the entanglement from the radiation density matrix into the environment and replaces the original state of the radiation with $|e_0\rangle\langle e_0|_R$; the interior code subspace could not possibly be protected under such a drastic replacement. But importantly, the SWAP channel is \emph{not} very complex\footnote{By the complexity of a quantum channel, we mean the minimum circuit complexity of the unitary $U_\cE$ in its isometric extension.} -- for every qubit in the radiation, we need one 2-qubit SWAP gate, and so the complexity of this channel should be $M\sim \log\,k$. 

Note, however, that by unitarity of quantum mechanics, if we treat the radiation as a subsystem and implement the erasure channel via a SWAP operator, this procedure does not actually destroy any information.
All it does is move where the black hole interior has been encoded in the larger system. 
This amounts to including the environment as a subsystem upon which we may try to reconstruct the interior.

\subsection*{Black hole exterior}
\begin{figure}
    \centering
    \includegraphics[height=5.6cm]{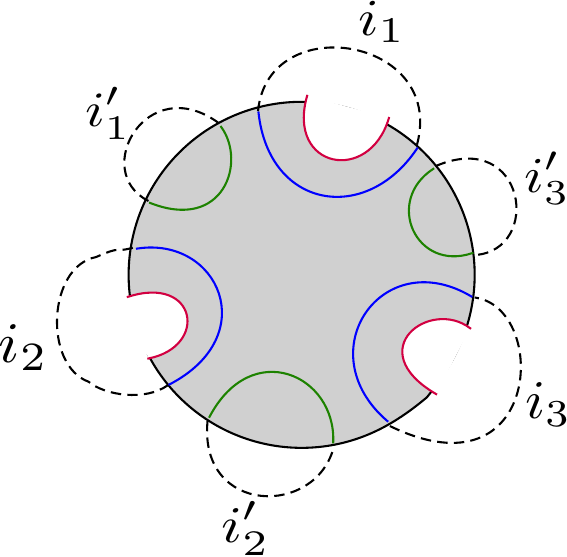} 
    \caption{\small{The connected gravitational contribution to the R\'enyi entropy of $\fr_i \cup \en$. Green and blue lines denote bulk contractions of the exterior and interior factors in the code subspace respectively. }}
    \label{fig:ext}
\end{figure}
We can also ask whether the operation $\cE$ can affect the black hole exterior. In this case, we need to compute the mutual information $I_{\Psi'}(\fr_e: \fr_i \cup \en)$. When the disconnected geometry dominates, the calculation is essentially the same as before and we get zero mutual information. When the connected geometry dominates, the calculation of $S(\fr_i\cup \fr_e \cup \en)$ is also the same as before. The main difference is in $\mathrm{Tr}\,{\rho'}^n_{\fr_i,\en}$ (Fig.~\ref{fig:ext}).
Instead of \eqref{eq:re-env-conn}, we get 
\beq
\mathrm{Tr}\,{\rho'}^n_{\fr_i,\en}\Big|_{\text{conn.}}\sim e^{(1-n)S_{BH}} \;\mathrm{Tr}\,\sigma^n_{R},
\eeq
where $(n-1)$ is small and positive. Note also that $\mathrm{Tr}\,{\rho'}^n_{\fr_e}$, which we need in place of \eqref{eq:disc-result3}, is always dominated by the disconnected geometry:
\beq
\mathrm{Tr}\,{\rho'}^n_{\fr_e}\Big|_{\text{disconn.}} =  d^{1-n}_e.
\eeq
Using these expressions, we can check that $I_{\Psi'}(\fr_e: \fr_i \cup \en)=0$, and thus the black hole exterior is never affected by quantum operations on the radiation, which is of course consistent with the fact that the exterior is not in the entanglement wedge of the radiation.

\subsection{Approximate isometry of the interior encoding}\label{sec:non-iso}

The theorems concerning standard quantum error correction  reviewed in Sec.~\ref{sec:QECreview} assumed an isometric embedding $V$ of the code Hilbert space in the physical Hilbert space.
However, the map $V\ket{i,i'}_\co = \ket{\Psi_{i,i'}}_\ph$ for the black hole interior is not actually an isometry.
This is because 
the enormous number of effective field theory states in the growing interior will eventually outnumber the states of the physical Hilbert space as measured by the black hole entropy.   
Thus, states that are orthogonal in the effective field theory cannot be orthogonal after embedding in the physical Hilbert space.
However, we might expect that 
for code subspaces which are small compared to the black hole Hilbert space, the interior encoding map $V$ should be an approximate  isometry \cite{Kim:2020cds}, and therefore the standard theorems should still be approximately valid and thus can be used to diagnose the existence of a recovery channel. 

We can use the gravitational path integral to demonstrate that this expectation is indeed true. The map $V$ can be made arbitrarily close to an isometry by appropriately tuning the code subspace relative to the black hole Hilbert space.
To see this, we will evaluate the Frobenius norm of the operator $P^2 - P$, where $P = V V^\dagger$ is the code subspace projector.
If $V$ were an isometry, $P$ would be a projector and we would have $P^2 - P = 0$.
So, if the Frobenius norm of this quantity is small, $V$ is approximately an isometry.

The first step is to normalize $P$ by requiring that 
\begin{equation}
    \frac{1}{\cN_P} \Tr P  = d ,
    \label{eq:projectornorm}
\end{equation}
where ${\cN_P}$ is a normalization and  $d$ is the code subspace dimension, as appropriate for normalizing a projector.  
Using $P = VV^\dagger$ and the definition of $V$ as above, the left hand side reduces to a combination of gravitational overlaps
\begin{equation}
    \frac{1}{\cN_P} \sum_{i,i'} \inner{\Psi_{i,i'}}{\Psi_{i,i'}}_\ph = d .
\end{equation}
These overlaps reduce to $\inner{\Psi_{i,i'}}{\Psi_{i,i'}}_\ph = e^{S_0} Z_1$ where we separated out the $e^{S_0}$ factor from the JT gravity path integral on the disc as before in \eqref{eq:norm-density-matrix}. 
Therefore, the normalization for $P$ is
\begin{equation}
    \cN_P = e^{S_0}Z_1 .
\end{equation}
As usual, we have ignored terms which are subleading in the genus expansion, which is justified as we have $S_0 \gg 1$.
This normalization essentially fixes the orthonormality of the code subspace states $\inner{i,i'}{j,j'}_\co = \delta_{ij}\delta_{i'j'}$.

We are now interested in the Frobenius norm 
\begin{equation}
    \| \widehat{P}^2 - \widehat{P} \|_F^2 \equiv \Tr (\widehat{P}^2-\widehat{P})^2 ,
\end{equation}
where we normalized $\widehat{P} \equiv P/\cN_P$ to have a proper projector (see Eq.~\ref{eq:projectornorm}).
The leading contribution to this norm in the genus expansion should arise from the fully disconnected contribution, where all gravitational boundaries are filled separately by disk geometries with an EOW brane.  In fact, an explicit calculations shows that these contributions cancel in the 
function of $\widehat{P}$ appearing in the Frobenius norm. 
To see this, we make the replacement $\inner{\Psi_{i,i'}}{\Psi_{j,j'}} = e^{S_0}Z_1 \delta_{ij}\delta_{i'j'}$.
The completely disconnected contributions are then
\begin{equation}
    \| \widehat{P}^2 - \widehat{P} \|_F^2 \Big|_{\text{total disconn.}} = \frac{d e^{4S_0} Z_1^4}{\cN_P^4}  - \frac{2d e^{3S_0} Z_1^3}{\cN_P^3}  + \frac{d e^{2S_0} Z_1^2}{\cN_P^2} = 0  .
\end{equation}
where the numerators were computed by direct evaluation of the relevant path integrals in the same manner as in the calculation of (\ref{eq:norm-density-matrix}).
Since this vanishes, we look for the next order contribution in the genus expansion. This comes from the path integral geometries in which precisely one pair of asymptotic boundaries is connected by a wormhole bounded on either side by EOW branes; we will call the JT path integral on this connected pair geometry $Z_2$. In $\Tr \widehat{P}^n$ this term appears from the overall path integral geometry with $n-1$ connected components; so the result will be suppressed by $e^{-S_0}$ compared to each term in the disconnected contribution.
Thus,
\begin{equation}
    \| \widehat{P}^2 - \widehat{P}\|_F^2 \Big|_{Z_2} = \frac{6d^2 e^{3S_0} Z_2Z_1^2}{\cN_P^4} - \frac{6d^2 e^{2S_0}Z_2Z_1}{\cN_P^3} + \frac{d^2e^{S_0}Z_2}{\cN_P^2} = d^2 e^{-S_0} \frac{Z_2}{Z_1^2} .
\end{equation}
The total Frobenius norm is therefore
\begin{equation}
    \| \widehat{P}^2 - \widehat{P}\|_F^2 = d^2 e^{-S_0} \frac{Z_2}{Z_1^2} + O(e^{-2S_0}) ,
\end{equation}
where  the $O(e^{-2S_0})$ term includes contributions with fewer disconnected components, and the genus expansion of each component.

This result tells us that by taking $d \ll e^{S_0/2}$, we can ensure that $\widehat{P}$ is exponentially close to a true projector in the physical Hilbert space.
Thus the assumption in \cite{Kim:2020cds} that the encoding map for the interior is close to an isometry holds true in the PSSY model of two-dimensional gravity.
To use the matrix ensemble language for JT gravity plus EOW branes \cite{Saad:2019lba}, we have shown that a typical member of the boundary ensemble implements an approximately isometric bulk-to-boundary encoding map $V$. So we are justified in drawing conclusions about the error correction properties of the typical theory by using the decoupling principle and similar standard isometric error correction theorems.

A more refined analysis is possible.
We could imagine deriving completely general Knill-Laflamme conditions for the existence of a recovery channel when the encoding map is arbitrarily far from an isometry.
In this most general case, $V$ is simply a linear map, not necessarily isometric, which must be injective to provide a faithful encoding. Injectivity certainly requires $d \ll e^{S_0}$.  However, whether this map is actually faithful depends on how it maps the code space into the actual physical Hilbert space.  If we are dealing with a theory of gravity like JT gravity that is dual to an ensemble of microscopic theories, faithfulness of the encoding $V$ requires that the measure in the ensemble should not be concentrated on theories for which $V$ is degenerate. In JT gravity, this is likely the case, as the EOW brane states which we labeled by $\alpha$ are dual to states that include random coefficients in the energy basis \cite{Penington:2019kki}, and are thus  highly unlikely to lead to degeneracies in $V$ for small $d$ compared to $e^{S_0}$.  However, despite this, it does not follow that the encoding map is necessarily an isometry, or even close to one, because the encoding vectors may have large overlaps in individual members of the ensemble.  If we could write generalized Knill-Laflamme conditions, along with a decoupling principle that applied to such non-isometric encodings, we could compute the relevant mutual information 
to diagnose the existence of a recovery channel for some error $\cE$ and given linear map $V$.
In this way, we could aim to derive entropic conditions for the existence of a recovery map in {\it every} member of the underlying ensemble for a given error, instead of showing its existence only for a typical member of the ensemble as we have done in this paper.

The non-isometric codes discussed in \cite{Akers:2021fut} may give some intuition for what the correct generalized Knill-Laflamme conditions should be.
In the language used in that work, in this paper  we are studying  a sort of state-independent sector of typical non-isometric codes where the dimension of the code subspace is small enough to ensure that more conventional error correction criteria are still approximately valid.

\section{Erasures and random errors}
\label{sec:ErasuresTypical}

The previous section developed general results about robustness against errors of the encoding of the black hole interior in  the Hawking radiation.   In this section we further investigate two kinds of error channels for which more detailed analyses are possible.  In the first class of errors, in which some of the physical bits in the radiation are simply erased, we are able to explicitly sum over all planar gravitational geometries (following the resolvent method of \cite{Penington:2019kki}) contributing to mutual information that diagnoses error correction.   In the second class consisting of typical errors, where a random unitary acts on the radiation and an auxiliary environment Hilbert space, we give a geometric formula for the mutual R\'enyi entropy in terms of gravitational path integrals, which when evaluated in JT gravity, reproduces the same coherent information bound obtained previously.

\subsection{Erasure errors}
Our analysis in the previous section was in the semi-classical approximation where we  restricted attention to replica-symmetric saddles. In this section, we will work with the full gravitational path integral, but in order to do so, we will need to specialize to a particular quantum channel: partial erasure of the radiation.
Suppose the error channel $\mathcal{E}_\text{erase}$ simply erases $t$ of the $M$ qubits in $\mathcal{H}_R$.
The channel applied to the maximally mixed state on $\mathcal{H}_R$ then yields a maximally mixed state of dimension $2^{M-t}$, since the partial trace over $t$ qubits of a maximally mixed state on $M$ qubits gives a maximally mixed state on $M-t$ qubits. 
A suitable isometric extension of $\mathcal{E}_\text{erase}$ is simply a partial analog of the SWAP operator discussed previously, where the  unitary $U_\cE$ exchanges $t$ ``clean'' environment qubits with $t$ radiation qubits. More precisely, by ``clean'' we mean that we take the environment to initially consist of $t$ qubits each in some fixed fiducial state $|0\rangle\langle 0|$, and the action of the channel swaps the state on $t$ of the radiation bath qubits with these $t$ environment qubits. Afterward, the remaining radiation entropy is $\log 2^{M-t}$ and the environment entropy is $\log 2^t$.
The coherent information is then
\begin{equation}
    I_c \left( \mathcal{E}_\text{erase} , \frac{1}{k} \mathbb{1} \right) = (M-2t) \log 2 .
\end{equation}

The  gravitational meaning of this result is surprising.
We have found that for, e.g., $t \approx M/2$, the coherent information vanishes or is at least very small in magnitude compared to a large black hole entropy.
This means that we can erase considerably more than half of the radiation and still reconstruct essentially all effective field theory operators in the black hole interior.
Hints of this robustness were seen in doubly holographic models of evaporation \cite{Balasubramanian:2020hfs}, where a multiboundary wormhole model was used to reach roughly the same conclusion.
But just how sharp is the transition between a correctable and non-correctable erasure error?
In other words, can we produce a more accurate criterion than our condition that the coherent information of the erasure channel must exceed minus the black hole entropy by ``a large amount'', as written in \eqref{CIcond}?
In order to study this transition more precisely, we must incorporate more gravitational geometries than just the disconnected and fully connected phases that we studied in Sec.~\ref{sec:qec-pssy}.

As discussed in Sec.~\ref{sec:qec-pssy}, in deriving our coherent information criterion we neglected the contribution from geometries breaking $\mathbb{Z}_n$ symmetry.  For erasure errors, we can do better by resumming all the planar geometries; this is a good approximation when the parameters $k, e^{S_0}, d_i$ etc. running in loops are large, and it will give us a sharper picture of the correctability transition for erasure errors.  We will perform the  planar resummation  by writing down a Schwinger-Dyson equation for a resolvent, following \cite{Penington:2019kki}.
The general idea of this method is to compute the entropy $S(\rho)$ of a density matrix $\rho$ indirectly by first computing a resolvent matrix $\cR(w) = [w\mathbb{1} - \rho]^{-1}$.
As is standard practice in, e.g., random matrix theory, we may obtain the eigenvalue density $D(w)$ for $\rho$ by computing the discontinuity of the resolvent across the real axis, $2\pi \mathrm{i} D(w) = \Tr [ \cR(w-\mathrm{i}\epsilon) - \cR(w+\mathrm{i}\epsilon) ]$ with $\epsilon \to 0$.
With $D(w)$ in hand, the entropy $S(\rho)$ is easily obtained by writing $S(\rho) = -\int \mathrm{d}w\; D(w) \; w \log w$.
We may similarly evaluate \textit{any} function of the eigenvalues of $\rho$ using $D(w)$; if we wish to compute $\Tr \rho^n$, for instance, we write $\Tr \rho^n = \int \mathrm{d}w \; D(w) w^n$.
The resolvent is particularly useful in summing planar diagrams for a density matrix $\rho$ that involves gravitational path integrals because there is a recursive Schwinger-Dyson equation that determines $\cR$ exactly in the planar approximation.

We are going to relax the Lewkowycz-Maldacena assumption of $\mathbb{Z}_n$ replica symmetry, and will replace it with three simplifying assumptions in order to keep the calculation tractable.
As we are most interested in subsystem correctability of $\cH_\inte$, we will assume that (i) we restrict ourselves to a code subspace where $\cH_{\ext}$ is trivial (i.e., one-dimensional). Next, as we noticed in Sec.~\ref{sec:qec-pssy}, part of the utility of the Lewkowycz-Maldacena assumption was that we were able to choose a particular contraction pattern for the bulk code subspace indices based on the geometry of the replica symmetry quotient manifold (see the discussion around Fig.~\ref{fig:conncode1}).  In relaxing the Lewkowycz-Maldacena assumption to incorporate replica symmetry breaking geometries, we want to continue choosing contraction patterns for the code subspace indices based only on the geometry of the replica manifold, so we assume that (ii) $\cH_{\inte}$ corresponds to a degree of freedom moving on or close to the EOW brane. Finally, the canonical ensemble involves integral equations relating the partition functions $Z_n$ and the density of states, so we will assume that (iii) we work in the microcanonical ensemble for the black hole, since the density of states in the microcanonical ensemble is flat in the microcanonical window and is therefore related to the microcanonical partition function by a simple algebraic equation; this will greatly simplify our Schwinger-Dyson equation for the resolvent.

Let us be more explicit about the implication of assumption (ii).
In products of gravitational microstate overlaps, if $\cH_\inte$ is a degree of freedom that is ``on or close to the EOW brane'', we can simply contract this index identically to the contraction pattern of the EOW branes.
For example, consider the overlap $\inner{\psi^{\alpha_1}_{i_1}}{\psi^{\beta_1}_{j_1}}\inner{\psi^{\alpha_2}_{i_2}}{\psi^{\beta_2}_{j_2}}_B$, where $i_1,j_1,i_2,j_2$ are the interior code subspace indices and $\alpha_1,\beta_1,\alpha_2,\beta_2$ are EOW brane indices.
Our assumption (ii) means that if we contract EOW brane indices to obtain $\delta_{\alpha_1\beta_1}\delta_{\alpha_2\beta_2}$, we require that the code subspace indices are contracted to yield $\delta_{i_1j_1}\delta_{i_2j_2}$.
Similarly, if we contract EOW brane indices as $\delta_{\alpha_1\beta_2}\delta_{\beta_1\alpha_2}$, we require that the code subspace contractions give $\delta_{i_1j_2}\delta_{j_1i_2}$.
This simplifies our analysis of replica symmetry breaking geometries considerably because it allows us to discard configurations where the EOW brane indices are contracted in a different pattern than the interior code subspace indices, a situation which was in principle possible in Sec.~\ref{sec:qec-pssy} but which did not actually arise due to the Lewkowycz-Maldacena assumption.
Here, we are relaxing the Lewkowycz-Maldacena assumption but we will keep one of its important implications by making assumption (ii) to keep the calculation tractable.

With these simplifications, we can  work out the R\'enyi mutual information $I^{(n)}_{\Psi'}(\fr_i:\en)$ in the relevant state
\begin{equation}
    \ket{\Psi'} = \frac{1}{\cN^{1/2}}\sum_{i=1}^{d_i} \ket{i}_{\fr_i} \otimes U_\cE (\ket{\Psi_i}_\ph \otimes \ket{e_0}_\en ) , \quad \cN = d_i e^{S_0} Z_1 ,
\end{equation}
where the normalization is the $d_e=1$ specialization of the one we had previously \eqref{eq:norm-density-matrix}.
Note that the R\'enyi mutual information  has the property that it vanishes if and only if the state factorizes. So, it is just as good as the mutual information $I_{\Psi'}(\fr_i : \en)$ as a diagnostic of  decoupling between the reference system and the environment. 

Let us first consider $\mathrm{Tr}\,\rho'^n_{\fr_i,\en}$. 
It is convenient to first write the matrix elements for the  reduced density matrix, each of which involves a gravitational overlap
\begin{equation}
    \bra{i,e_m} \rho'_{\fr_i,\en} \ket{j,e_n} = \frac{1}{k \cN} \sum_{\alpha,\beta,\gamma=1}^k  \bra{\gamma,e_m} U_\cE \ket{\beta,e_0} \inner{\psi^\alpha_i}{\psi^\beta_j} \bra{\alpha,e_0} U_\cE^\dagger \ket{\gamma,e_n}.
\label{eq:Ri-env-element}
\end{equation}
We can sum over planar geometries by introducing the resolvent matrix $\cR$:
\beq
\cR(w) \equiv \frac{1}{(w \mathbb{1} - \rho'_{\fr_i,\en})}= \frac{1}{w}\mathbb{1}+\frac{1}{w^2}\rho'_{\fr_i,\en}+\frac{1}{w^3}\rho'^2_{\fr_i,\en}+\cdots,
\label{eq:resolvent-expansion}
\eeq
where $w$ is a complex number and $\cR$ has dimension $\dim (\cH_\inte \otimes \cH_\en)$. 
We may represent this infinite sum pictorially using the same gravitational boundary conditions and tensor network constructions as in Fig.~\ref{fig:f}: 
\beq
\includegraphics[height=0.8cm]{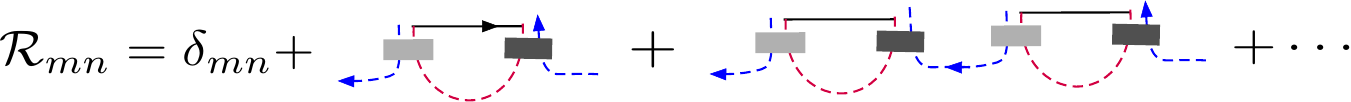}
\label{eq:resolvent-bc-pic}
 \eeq
Here the black lines denote asymptotic boundaries and come with a factor of $(k\cN)^{-1}$, dashed blue lines denote environment indices, and dashed red lines denote radiation indices; we have suppressed the $\fr_i$ index for simplicity. Further, $U_\cE$ are denoted with light gray boxes and $U_\cE^\dagger$ are denoted with dark gray boxes. The $n$th term additionally also comes with the appropriate power $w^{-n}$ which has been left implicit in the figure.
 
 As an example of how \eqref{eq:resolvent-expansion} is translated to \eqref{eq:resolvent-bc-pic}, take the first non-trivial term in \eqref{eq:resolvent-bc-pic}, which is pictorially representing the matrix element \eqref{eq:Ri-env-element}.
 The gravitational boundary condition is $\inner{\psi_i^\alpha}{\psi^\beta_j}_B$, represented by the solid black line oriented from ket to bra.
 Implicit at the left and right endpoints of this condition are the $(\beta,j)$ and $(\alpha,i)$ indices, respectively.
 The left endpoint with radiation index $\beta$ feeds into the light gray $U_\cE$ operator, along with a hanging environment leg that is implicitly contracted with $e_0$.
 The outgoing environment leg of $U_\cE$ is $e_m$, while the outgoing radiation leg is $\gamma$, and this radiation index is matched to the incoming radiation index of the dark gray $U_\cE^\dagger$ operator.
 The incoming environment leg in $U_\cE^\dagger$ is $e_n$, while the outgoing radiation leg is $\alpha$, matched to the EOW brane index at the right endpoint of the black line.
 Finally, the outgoing radiation leg on $U_\cE^\dagger$ is implicitly contracted with $e_0$.
 We have suppressed the $\fr_i$ indices since these would just add an extra incoming and outgoing dashed line attaching to the endpoints of the black line, but not interacting with the $U_\cE$ or $U_\cE^\dagger$ operators.
 The missing factors of $w$ and $k\cN$ are implicit as mentioned under \eqref{eq:resolvent-bc-pic}.

 Evaluating \eqref{eq:resolvent-bc-pic} exactly would be very difficult if we were to allow arbitrary gravitational geometries filling in the boundary conditions.
 However, in the planar approximation, there is an interesting simplification, as exploited in \cite{Penington:2019kki}. 
 If we only consider planar gravitational geometries, namely those whose index loops and bulk gravitational components do not cross each other and do not include higher genus contributions, there is a recursive Schwinger-Dyson equation for this resolvent where the right hand side of \eqref{eq:resolvent-bc-pic} can be written pictorially as
  \beq
\includegraphics[height=2.5cm]{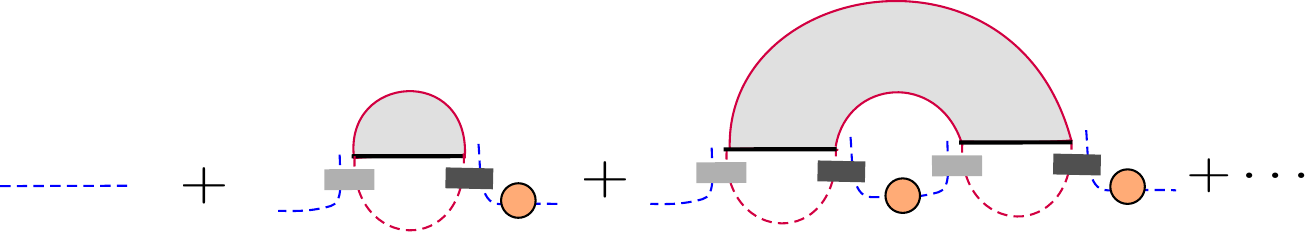},
\label{eq:SD-pic}
 \eeq
where the shaded regions denote bulk Euclidean spacetimes filling in the asymptotic boundary conditions with the solid red lines denoting EOW branes, and the orange blobs stand for insertions of the resolvent.
The logic behind why this sort of recursion captures all the planar contributions is as follows.
The first nontrivial term in \eqref{eq:SD-pic} captures all planar geometries in which the first asymptotic boundary is not connected through the bulk to any other asymptotic boundary.
Similarly, the second nontrivial term captures all planar geometries in which the first asymptotic boundary is connected through the bulk to just one other asymptotic boundary, but this second boundary does not necessarily have to be adjacent to the first, and it also does not necessarily have to be the last asymptotic boundary on the right.
To capture the situations in which they are not adjacent, the resolvent (orange blob) is inserted between the two boundaries, and to capture the situations in which the second boundary is not also the last boundary on the right, the resolvent also is inserted to the right of the second boundary.
These two resolvent insertions are independent because any connection through the bulk between them would render the geometry non-planar.
Iterating this sequence, and noting that every planar geometry can be put into exactly one such class, we arrive at the recursion relation \eqref{eq:SD-pic}.

We can translate \eqref{eq:SD-pic} back to an equation by recalling several facts. 
First, $Z_m$ is the JT gravity path integral on a disk with $m$ EOW branes, and clearly the $m^{\text{th}}$ nontrivial term on the right hand side of \eqref{eq:SD-pic} carries a factor of $Z_m$.
Second, $m$ explicit gravitational boundary conditions come with a normalization of $(k\cN)^{-m}$ as described below \eqref{eq:resolvent-bc-pic}.
Also, dashed blue environment index lines come with a factor of $w^{-1}$, unless they have a resolvent insertion, in which case this factor is omitted since there is no factor of $w$ on the left hand side of \eqref{eq:resolvent-expansion}.
Because of this, from the form of \eqref{eq:SD-pic}, we see that every term on the right hand side is proportional to $w^{-1}$.
By our assumption (ii), the code subspace indices are contracted along the EOW branes, which means the $\fr_i$ indices of each resolvent insertion (except the last one on the right) in each term in \eqref{eq:SD-pic} are contracted together, yielding a trace $\Tr_{\fr_i} \cR$.

Defining the $R \cup \en$ density matrix $\sigma$ again as
\beq
 \sigma = U_\cE \left(\frac{1}{k}\sum_{\alpha=1}^k|\alpha\rangle\langle \alpha|_R \otimes |e_0\rangle\langle e_0|_{\en}\right)U_\cE^{\dagger} ,
\eeq
we see that because the resolvents are interspersed only on the environment factors, they multiply between the $\sigma$ factors to give $[\Tr_\en (\sigma . \Tr_{\fr_i} \cR)]^{m-1}$.  Here $m-1$ appears instead of $m$ since the last factor of $\sigma$ is generated by the leftmost $U_\cE$ and the rightmost $U_\cE^\dagger$, and is therefore not included in the environment traces generated by the adjacent EOW brane contractions, as can be seen from studying \eqref{eq:conn-EOW-structure} and \eqref{eq:f-tensor}.
This last factor of $\sigma$ instead has a radiation trace from the single large EOW brane index loop in $Z_m$.
Tacking on the last factor of $\cR$ on the right, which multiplies the environment factor from the right, we finally find
 \beq
 \cR 
 =\frac{1}{w}\mathbb{1}_{\fr_i,\en}+ \frac{1}{w} \sum_{m=1}^{\infty}\frac{Z_m}{\cN^m}\mathrm{Tr}_R\{\sigma. [\mathrm{Tr}_{\fr_i, \en}(\sigma.\cR)]^{m-1}\}.\cR,
 \label{eq:SD-eqn}
 \eeq
 The dot notation for products means that we multiply operators which may act on different Hilbert spaces by appending an identity operator to the extra factors.
 Note the factor of $k^{-m}$ from the normalization was used to construct the $m$ factors of $\sigma$.

For a general quantum channel, solving \eqref{eq:SD-eqn} seems involved, although not impossible. But our restriction to the partial SWAP channel simplifies things quite a bit. For this channel, we have the factorization
 \beq
 \sigma = \sigma_R \otimes \sigma_\en, \quad \sigma_\en = \frac{1}{2^t}\mathbb{1}_\en, \quad \sigma_R = \frac{1}{2^{M-t}}\mathbb{1}_{M-t}\otimes (|e_0\rangle\langle e_0|)^{\otimes t} ,
 \eeq
 so we can rewrite the Schwinger-Dyson equation \eqref{eq:SD-eqn} as
 \beqn
 w\cR &=&\mathbb{1}_{\fr_i,\en}+\sum_{m=1}^{\infty}\frac{Z_m}{\cN^m}\mathrm{Tr}_R(\sigma_R^m) \left(\mathrm{Tr}_{\fr_i, \en}(\sigma_\en.\cR)\right)^{m-1} \sigma_\en.\cR,\nonumber\\
 &=&\mathbb{1}_{\fr_i,\en}+\frac{1}{2^t}\sum_{m=1}^{\infty}\frac{Z_m}{\cN^m2^{M(m-1)}} \left(\mathrm{Tr}_{\fr_i, \en}(\cR)\right)^{m-1} \cR
 \eeqn
In fact, the first line also holds for any channel $U_\cE$ which does not generate any mutual information when acting on the maximally mixed radiation state times the fiducial environment state $\ket{e_0}\bra{e_0}_\en$; in other words when $I_{\sigma}(R : \en) = 0$, as this implies $\sigma = \sigma_R \otimes \sigma_\en$.
Note that this is a different mutual information from $I_{\Psi'}(\fr_i : \en)$, which we use to diagnose correctability of $U_\cE$. To go further, we need expressions for the partition functions $Z_n$, which are found in \cite{Penington:2019kki}.
We have the integral expression
\beq
 Z_n = \int_0^{\infty}ds\,\rho(s)[y(s)]^n,
 \eeq
 where
 \beq
 \rho(s) = \frac{s}{2\pi^2}\sinh(2\pi s),\;\;\;y(s)=e^{-\frac{\beta}{2}s^2}2^{1-2\mu}|\Gamma(\mu-\frac{1}{2}+is)|^2,
 \eeq
 after which the resolvent sum becomes geometric:
 \beq
 w\cR=\mathbb{1}_{\fr_i,\en}+\frac{1}{2^t}\int_0^{\infty}ds\frac{1}{\cN }\rho(s)y(s)\frac{1}{1-\frac{y(s) \mathrm{Tr}_{\fr_i,\en}(\cR)}{\cN 2^{M}}}\cR .
 \eeq
 
The integral equation relating the density of states $\rho(s)$ to the JT path integral $Z_n$ is complicated due to the dependence of the integrand on the energy $s$.
By employing assumption (iii), we may work instead in the microcanonical ensemble where we define the gravitational path integral with boundary conditions that will fix the energy $s$ in a small window $\Delta s$.
This choice will change nothing in the above Schwinger-Dyson analysis except the explicit form of $Z_n$, which is now \textit{algebraically} related to a flat microcanonical density of states in the energy window $[s,s+\Delta s]$ with height $e^{S_0} \rho(s)$.
Specifically, we have the microcanonical entropy $\textbf{S}$ and microcanonical partition functions $\textbf{Z}_n$ defined by
\begin{equation}
    e^{\textbf{S}} = e^{S_0}\rho(s) \Delta s , \quad \textbf{Z}_n = e^{S_0}\rho(s) y(s)^n \Delta s = e^{\textbf{S}} y(s)^n .
\end{equation}
These relations simplify the above integral equation as we replace $e^{S_0} Z_n$ by $\textbf{Z}_n$ and $\int \mathrm{d}s$ by $\Delta s$, and after taking a trace over $\fr_i \cup \en$ the microcanonical Schwinger-Dyson equation for $\Tr \cR$ is
  \begin{figure}
     \centering
       \includegraphics[height=7.5cm]{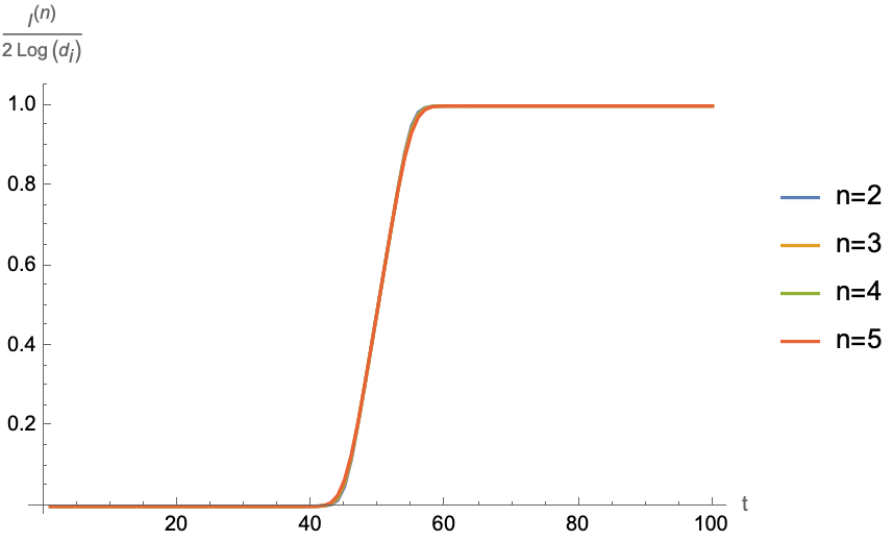}    
     \caption{\small{The R\'enyi mutual information from $n=2$ to $n=5$ as a function of the number of qubits $t$ in the environment. We have taken $d_i=2^{10},\;e^{\cS}=2^{100}$. We see that the phase transition happens precisely near $t\sim \frac{\cS-\log\,d_i}{2 \log 2}= 45.$ The curves overlap substantially, except near the phase transition.}}
     \label{fig:Renyi}
 \end{figure}
  \beq
 w\,X= d_i2^t+\frac{1}{d_i2^t}\frac{X}{\left(1-\frac{ X}{d_ie^{\textbf{S}} 2^{M}}\right)}, \quad X=\mathrm{Tr}_{\fr_i,\en}\;\cR.
 \eeq

This equation can be solved to obtain
 \beq
 \mathrm{Tr}_{\fr_i,\en}\,\cR^{(\fr_i,\en)} = \frac{d_ie^{\cS}}{2}-\frac{\left(2^{-t}e^{\cS}-d_i2^t\right)}{2w}-\frac{1}{2}\sqrt{\left(d_ie^{\cS}-\frac{\left(2^{-t}e^{\cS}-d_i2^t\right)}{w}\right)^2-4\frac{d_i^2e^{\cS}2^t}{w}},
 \eeq
 where we have added a superscript $(\fr_i,\en)$ to $\cR$ to denote the fact that this is the resolvent for the density matrix $\rho'_{\fr_i,\en}$, and we have defined $e^{\cS}=2^Me^{\textbf{S}}$.  Similarly, we can also obtain the resolvent $\cR^{(\en)}$ for $\rho'_{\en}$:
  \beq
 \mathrm{Tr}_{\en}\,\cR^{(\en)} = \frac{d_ie^{\cS}}{2}-\frac{\left(2^{-t}d_ie^{\cS}-2^t\right)}{2w}-\frac{1}{2}\sqrt{\left(d_ie^{\cS}-\frac{\left(2^{-t}d_ie^{\cS}-2^t\right)}{w}\right)^2-4\frac{d_ie^{\cS}2^t}{w}}.
 \eeq
By expanding these resolvents in powers of $\frac{1}{w}$, we can  compute all the R\'enyi entropies. We show a comparison of the R\'enyi mutual information as a function of $t$, the number of environment qubits, between several R\'enyi indices in Fig.~\ref{fig:Renyi}. From these plots we indeed see that $\mathbb{Z}_n$-symmetry breaking contributions only become important close to the phase transition and smooth out a relatively sharp transition between the fully disconnected and the fully connected geometries. 
These two $\mathbb{Z}_n$-symmetric contributions were evaluated in general in Sec.~\ref{sec:qec-pssy} and together create a step function shape for the R\'enyi mutual information, so we interpret the near-step function shape of the full planar resummation as just a small correction to these contributions, at least for erasure errors.

Note also that the transition happens near, but slightly before  
\begin{equation} 
2t \sim \frac{\cS}{\log 2} - \log_2 d_i , 
\label{eq:gravity-singleton}
\end{equation} 
where $\cS$ is the logarithm of the total physical Hilbert space dimension (microcanonical JT ensemble with entropy $\textbf{S}$ plus the radiation bath with entropy $M \log 2$) and $t \log 2$ is the logarithm of the environment Hilbert space dimension.  
This is precisely the upper limit from the quantum Singleton bound \eqref{eq:singleton-bound} on the number of qubits that can be erased while leaving the code subspace protected.
More specifically, in \eqref{eq:singleton-bound} we called $\cS$ the number of qubits in the physical Hilbert space, so we must replace it here with $\cS / \log 2$, and we had a base-2 logarithm of the code subspace dimension, both of which appear in \eqref{eq:gravity-singleton}.
The R\'enyi mutual information starts to become $O(1)$ (as opposed to being exponentially small) when $t$ is only an $O(1)$ number less than the greatest number of qubit erasures that any code can protect against!
Thus, the JT gravity model comes within an $O(1)$ number of qubits of saturating the quantum Singleton bound.
It would be interesting to characterize the very slightly suboptimal performance of gravity; perhaps non-planar contributions are important to account for in this analysis.

\subsection{Typical errors}
\label{sec:randomerrors}

In this section we will consider how well the black hole interior is protected against the action of typical random errors on the radiation bath. 
In more detail, suppose that the Stinespring dilation of $\mathcal{E}$ involves a unitary operator $U_\cE$ acting on a Hilbert space of dimension $\dim (\cH_R \otimes \cH_\en) = k\ell$, where $\ell$ is the dimension of the environment factor and $k$ controls the dimension of the radiation subspace that is entangled with the black hole microstates.
Then, by a typical/random error, we mean one where we can take $U_\cE$ to be a random unitary transformation drawn from the Haar measure on the group $U(k\ell)$. 

We will perform this random error calculation in two ways.
First, we will directly Haar average the $f$-tensor \eqref{eq:f-tensor} and re-compute the mutual information $I_{\Psi'}(\fr_i : \fr_e \cup \en)$.
Second, we will take the Haar average of the coherent information $I_c(\cE,\mathbb{1}_R/k)$ and compare its value with our correctability criterion \eqref{CIcond}.
A priori, these two approaches need not give the same result, because averaging the $f$-tensor first may affect which gravitational geometries dominate the path integral, whereas in averaging the coherent information we have assumed that the replica symmetric contributions are the relevant ones.
However, we will see that an interesting simplification occurs when we consider the two limits $1 \ll k \ll \ell$ and $1 \ll \ell \ll k$, and the two approaches in fact agree.

\subsubsection{$f$-Tensor}
\label{sec:f-tensor-haar}

We begin by directly taking the Haar average of \eqref{eq:f-tensor} and evaluating the resulting gravitational path integrals which appear in the mutual information (\ref{eq:renyimutualinformation}) between the interior and the union of the exterior and the environment. 
This in turn involves the R\'enyi entropies of various combinations of the black hole interior, exterior and the environment (see Eqs.~\ref{tr1},\ref{tr2},\ref{tr3}).   
We will first average the error over the unitary group and then carry out the gravitational path integral computing the microstate overlaps.
As we are most interested in the regimes where the environment is either large ($\ell \gg k \gg 1$) or small ($1 \ll \ell \ll k$) compared to the radiation Hilbert space with both large compared to 1, we will only retain the leading terms in $1/k$ or $1/\ell$.
The Haar average of the $f$-tensor is then:
\begin{align}
\begin{split}
\bar{f}_{\alpha_1,\beta_1,\cdots \alpha_n,\beta_n}
&=\int_{U(k\ell)} dU_\cE  f_{\alpha_1,\beta_1,\cdots \alpha_n,\beta_n}(U_\cE) \\
& = \sum_{m_1,\cdots, m_n}\sum_{\gamma_1,\cdots, \gamma_n} \int_{U(k\ell)} dU_\cE  (U^{\dagger})^{\alpha_1,e_0}_{\gamma_1,e_{m_2}}  U^{\gamma_1,e_{m_1}}_{\beta_1,e_0}  \cdots (U^{\dagger})^{\alpha_n,e_0}_{\gamma_n, e_{m_1}}   U^{\gamma_n,e_{m_n}}_{\beta_n,e_0}
\label{eq:HaarAverageU}
\\
& = (\ell^{1-n} + \cdots) (\delta_{\alpha_1\beta_1} \delta_{\alpha_2\beta_2} \ldots  \delta_{\alpha_n\beta_n}) + (k^{1-n} + \cdots ) (\delta_{\alpha_1\beta_2} \delta_{\alpha_2\beta_3} \ldots  \delta_{\alpha_n\beta_1} )   + \cdots ,
\end{split}
\end{align}
where we have defined the matrix elements 
\begin{equation} 
U^{\beta,e_m}_{\alpha,e_n} \equiv \bra{\beta,e_m}U_\cE \ket{\alpha,e_n} , \quad  (U^\dagger)^{\beta,e_m}_{\alpha,e_n} \equiv \bra{\beta,e_m}U_\cE^\dagger \ket{\alpha,e_n} .
\end{equation}
We compute the integrals in the second line in Appendix~\ref{sec:haar} by applying standard results concerning Haar integration of products of unitary matrices.  
The first term dominates when $1 \ll \ell \ll k$, while the second term dominates when $1 \ll k \ll \ell$.

Remarkably, the first term in (\ref{eq:HaarAverageU}) has the same microstate index structure as the product of disconnected path integrals (\ref{eq:DiscOverlap}) for the  gravitational overlaps in the R\'enyi mutual informations (\ref{tr1}), (\ref{tr2}), and (\ref{tr3}). The second term has the same microstate index structure as the fully connected diagrams for computing the same product of overlaps, shown for instance in (Fig.~\ref{fig:conncode2} and Eq.~\ref{eq:conn-EOW-structure}). This means that averaging over $U_{\cE}$, at least in the two limits $1 \ll \ell \ll k$ and $1 \ll k \ll \ell$, reinforces the two replica-symmetric, disconnected and connected geometries respectively in the gravitational path integral.   Due to this simplification, we see that the relevant gravitational geometries that will enter in the calculation of the R\'enyi mutual information $I_{\Psi'}^{(n)}(\fr_i : \fr_e \cup \en)$ are precisely the ones we considered already in Sec.~\ref{sec:qec-pssy}. In the $n\to 1^+$ limit, the overlaps in (\ref{tr1}), (\ref{tr2}), and (\ref{tr3}) involve different propagators for the matter fields and thus lead to different powers of $d_i$ and $d_e$ (see discussion below Eq.~\ref{eq:conn-EOW-structure}), but in a manner identical to Sec..~\ref{sec:qec-pssy}. 
Therefore, the resulting mutual information will be identical, as we will now confirm explicitly.

Using our result \eqref{eq:HaarAverageU} for the Haar-averaged $f$ tensor in (\ref{tr1}), (\ref{tr2}), and (\ref{tr3}),  along with the gravitational overlaps from the disconnected saddle-points (\ref{eq:DiscOverlap}), we can find  the leading  disconnected contribution to the R\'enyi entropies.
The contraction of $\bar{f}$ with the tensor on the radiation Hilbert space coming from the gravitational overlaps gives
\begin{equation}
\frac{1}{k^n}
\sum_{\alpha_1...\alpha_n}
\sum_{\beta_1...\beta_n}
\left(
\delta_{\alpha_1\beta_1} \delta_{\alpha_2\beta_2} \cdots \delta_{\alpha_n\beta_n} 
\right) \bar{f}_{\alpha_1\beta_1 \cdots \alpha_n\beta_n}
= \ell^{1-n} + k^{2(1-n)} + \cdots
\label{eq: g-f-contraction1}
\end{equation}
Carrying out the remaining sums over the internal and external code space indices gives
\begin{eqnarray}
\mathrm{Tr}\,{\rho'}^n_{\fr_i,\fr_e,\en} 
&=&
d_i^{1-n}  d_e^{1-n}  \, 
\left[ 
\ell^{1-n} + k^{2(1-n)}
+ \cdots
\right]   
\label{eq:TrIEDisc}
\\
\mathrm{Tr}\,{\rho'}^n_{\fr_e,\en}
&=&
d_e^{1-n} \,
\left[ 
\ell^{1-n} + k^{2(1-n)}
\right]
\label{eq:TrEDisc}
\\
\mathrm{Tr}\,{\rho'}^n_{\fr_i} 
&=&
d_i^{1-n} \, .
\label{eq:TriDisc}
\end{eqnarray}
The terms in the square brackets give an explicit evaluation of the $\Tr{\sigma_{\rm env}^n}$ factors in (\ref{eq:disc-result1})-(\ref{eq:disc-result3}) for the disconnected gravitational contribution to a general error. We can now evaluate the $n$-th R\'enyi mutual information (\ref{eq:renyimutualinformation}) between the interior of the black hole and the exterior of the black hole plus the environment  in the disconnected phase of the gravitational path integral. We find
\begin{equation}
    I^{(n)}(\fr_i: \fr_e \cup \en) = 
    \frac{1}{1-n} \left( \log \mathrm{Tr}\,{\rho'}^n_{\fr_i} +  
    \log \mathrm{Tr}\,{\rho'}^n_{\fr_e,\en}
    -
    \log \mathrm{Tr}\,{\rho'}^n_{\fr_i,\fr_e,\en} 
    \right)
    =
    0 \, ,
\end{equation}
confirming  (\ref{eq:disconnMI}) for random error channels.

Similarly, we can combine the Haar-averaged f-tensor with the gravitational overlaps from the connected saddlepoints in \eqref{eq:conn-EOW-structure}, along with its analogs for the other mutual informations, to computed the leading connected contribution to the R\'enyi entropies (\ref{tr1}), (\ref{tr2}), and (\ref{tr3}).  The contraction of $\bar{f}$ with the tensor on the radiation Hilbert space coming from the connected gravitational overlaps gives
\begin{equation}
\frac{1}{k^n}
\sum_{\alpha_1...\alpha_n}
\sum_{\beta_1...\beta_n}
\left(\delta_{\alpha_1\beta_2} \delta_{\alpha_2\beta_3} \cdots \delta_{\alpha_n\beta_1} \right) \bar{f}_{\alpha_1\beta_1 \cdots \alpha_n\beta_n}
=
(k\ell)^{1-n} + k^{1-n} + \cdots
\label{eq: g-f-contraction2}
\end{equation}
Carrying out the remaining sums over the internal and external code space indices (in the $n\to 1^+$ limit) gives
\begin{eqnarray}
\mathrm{Tr}\,{\rho'}^n_{\fr_i,\fr_e,\en} 
&=&    \frac{e^{S_0 (1-n)} Z_n}{Z_1^n}
\, d_e^{1-n} \,
\left[ 
(k \ell)^{1-n} + k^{1-n}
\cdots \right]
\label{eq:TrIEConn}
\\
\mathrm{Tr}\,{\rho'}^n_{\fr_e,\en}
&=&\frac{e^{S_0 (1-n)} \, Z_n}{Z_1^n} \,  d_i^{1-n}   d_e^{1-n} \, \left[ 
(k \ell)^{1-n} + k^{1-n}
+  
\cdots \right]
\label{eq:TrEConn}
\\
\mathrm{Tr}\,{\rho'}^n_{\fr_i} 
&=& 
\frac{e^{S_0 (1-n)} Z_n}{Z_1^n} \, k^{1-2n} \ell^{-n}
\label{eq:TriConn}
\end{eqnarray}
The quantities $S_0$, $Z_n$ and $Z_1$ are discussed below (\ref{eq:conn-EOW-structure}). The terms in the square brackets give an explicit evaluation of the $\Tr{\sigma_{\rm R}^n}$ factors in (\ref{eq:re-env-conn1})-(\ref{eq:re-env-conn}) for the connected gravitational contribution to a general error.

We can now evaluate the R\'enyi mutual information (\ref{eq:renyimutualinformation}) between the interior and  exterior  plus the environment in the connected phase of the gravitational path integral in the $n\to 1$ limit.  In evaluating this mutual information we will continue to use  the disconnected expression for the $\mathrm{Tr}\,{\rho'}^n_{\fr_i}$ (\ref{eq:TriDisc}) because it is dominant for any choice of the error rank $\ell$ as the connected contribution (\ref{eq:TriConn}) is suppressed in powers of $k$.  We get 
\beq
\begin{split}
I(\fr_i: \fr_e \cup \en) = 
   \lim_{n\to 1} \frac{1}{1-n} \left( \log \mathrm{Tr}\,{\rho'}^n_{\fr_i} +  
    \log \mathrm{Tr}\,{\rho'}^n_{\fr_e,\en}
    -
    \log \mathrm{Tr}\,{\rho'}^n_{\fr_i,\fr_e,\en} 
    \right)
    = 2 \log d_i
\end{split}
\eeq
confirming the general result (\ref{eq:connMI}) for random error channels.

Finally we can check the condition for the connected phase to dominate the R\'enyi entropies and hence the mutual information between the black hole interior and the exterior plus the environment.  As discussed above,   $\mathrm{Tr}\,{\rho'}^n_{\fr_i}$ is always dominated by the disconnected component because the connected part (\ref{eq:TriConn}) is suppressed.   A little algebra shows that the connected contribution to $\mathrm{Tr}\,{\rho'}^n_{\fr_e,\en}$ in (\ref{eq:TrEConn}) dominates the disconnected contribution in (\ref{eq:TrEDisc}) when
\begin{equation}
    \frac{\left(\frac{\ell}{k} \right)^{n-1} + k^{n-1}}{1 + \ell^{n-1}}
        \approx 
    \left( \frac{k}{\ell} \right)^{n-1} 
        \ll 
    e^{(1-n)( S_{BH} + \log d_i )}
\end{equation}
where $e^{S_0(1-n)} Z_n / Z_1^n \sim e^{(1-n) S_{BH}}$ in the $n\to 1^+$ limit, as discussed below (\ref{eq:re-env-conn}).  The approximate expression is valid when $\ell \sim O(k^2)$ or smaller. Similarly the connected contribution to $\mathrm{Tr}\,{\rho'}^n_{\fr_i,\fr_e,\en}$  in (\ref{eq:TrIEConn}) dominates the disconnected contribution in (\ref{eq:TrIEDisc}) when
\begin{equation}
    \frac{\left(\frac{\ell}{k} \right)^{n-1} + k^{n-1}}{1 + \ell^{n-1}}
        \approx 
    \left( \frac{k}{\ell} \right)^{n-1} 
        \ll 
    e^{(1-n)( S_{BH} - \log d_i )}
    \label{eq:secondcondition}
\end{equation}
The approximate expression is again valid when $\ell \sim O(k^2)$ or smaller.  This second condition is satisfied for smaller $\ell$ than the first condition.

It can be checked that the mutual information is nonzero if the connected phase dominates just $\mathrm{Tr}\,{\rho'}^n_{\fr_i,\fr_e,\en}$. The condition (\ref{eq:secondcondition}) is satisfied when
\begin{equation}
    \ell \gg \frac{k}{d_i} e^{S_{BH}} \gg k
\label{eq:thirdcondition}
\end{equation}
giving a precise condition on the rank of the error channel required to give an error that cannot be corrected.

\subsubsection{Coherent information criterion}
\label{sec:CIcond-haar}

Alternatively we can diagnose correctability of typical errors by directly computing the Haar average of the coherent information criterion in (\ref{eq:coherentinfo}).
A priori, this is a separate calculation from the one we discussed above, where we took the Haar average of the $f$-tensor before computing the gravitational path integral.
Here, we will take the general correctability criterion, which in particular is nonlinear in the error channel $U_\cE$, and perform the Haar average.
However, in the case of the $f$-tensor average, because we found that the microstate index structures precisely matched the two relevant ones coming from the disconnected and fully connected gravitational path integral geometries, we ought to obtain the same answer by simply averaging the coherent information.
As we will now see, this is indeed what occurs, and we obtain the typical error condition \eqref{eq:secondcondition} in a more straightforward manner, without discussing gravitational path integrals.

The coherent information is given by the difference of two entropies, $S(\sigma_R)$ and $S(\sigma_\en)$, where we recall that $\sigma_R$ and $\sigma_\en$ are partial traces of $\sigma$
\begin{equation}
    \sigma = U_\cE \left( \frac{1}{k} \mathbb{1}_R \otimes \ket{e_0}\bra{e_0}_\en \right) U_\cE^\dagger .
\end{equation}
over the complementary factor (see (\ref{eq:sigmaenv}) and (\ref{eq:sigmaR})).
Choosing explicit bases $\ket{\alpha}_R$ with $\alpha = 1, \dots , k$ and $\ket{e_m}_\en$ with $m = 1, \dots , \ell$ for the radiation and environment respectively, the reduced density matrices are
\begin{equation}
\begin{split}
    \sigma_R & = \frac{1}{k} \sum_{\alpha,m} \bra{e_m} U_\cE \ket{\alpha, e_0} \bra{\alpha, e_0} U_\cE^\dagger \ket{e_m} , \\
    \sigma_\en & = \frac{1}{k} \sum_{\alpha,\beta} \bra{\beta} U_\cE \ket{\alpha, e_0} \bra{\alpha, e_0} U_\cE^\dagger \ket{\beta} .
\end{split}
\end{equation}
We can compute the Haar averaged entropies by using the replica trick just as we did with the gravitational path integral.
Note that here we are not actually performing any gravitational path integrals; we are simply using the replica expressions for entropies and computing their Haar averages in order to find the typical value of the coherent information for a random channel.
The $n^{\text{th}}$ R\'enyi coherent information with respect to the maximally mixed state is given in terms of the $n^{\text{th}}$ R\'enyi entropies by
\begin{equation}
    I^{(n)}_c \left( \mathcal{E}, \frac{1}{k} \mathbb{1}_R \right) =  S^{(n)}(\sigma_R) - S^{(n)}(\sigma_\en) .
\end{equation}
The necessary traces of powers of the above density matrices are given by
\begin{equation}
\begin{split}
    \Tr \sigma_R^n & = \frac{1}{k^n} \sum_{\alpha_1, \cdots , \alpha_n} \sum_{\beta_1, \cdots , \beta_n} \sum_{m_1, \cdots , m_n} U^{\beta_1,e_{m_1}}_{\alpha_1,e_0} (U^\dagger)^{\alpha_1,e_0}_{\beta_2, e_{m_1}} \cdots U^{\beta_n,e_{m_n}}_{\alpha_n,e_0} (U^\dagger)^{\alpha_n,e_0}_{\beta_1, e_{m_n}} , \\
    \Tr \sigma_\en^n & = \frac{1}{k^n} \sum_{\alpha_1, \cdots , \alpha_n} \sum_{\beta_1,\cdots \beta_n} \sum_{m_1, \cdots , m_n} U^{\beta_1,e_{m_1}}_{\alpha_1,e_0} (U^\dagger)^{\alpha_1,e_0}_{\beta_1,e_{m_2}} \cdots U^{\beta_n,e_{m_n}}_{\alpha_n,e_0} (U^\dagger)^{\alpha_n,e_0}_{\beta_n,e_{m_1}} ,
\end{split}
\end{equation}

The Haar averages $\overline{\Tr \sigma_R^n}$ and $\overline{\Tr \sigma_\en^n}$ can be evaluated in terms of Weingarten functions and  combinations of Kronecker delta functions, as explained in Appendix~\ref{sec:haar}.
The results are 
\begin{equation}
\begin{split}
    \overline{\Tr \sigma_R^n} & = \frac{1}{k^{n-1}} , \\
    \overline{\Tr \sigma_\en^n} & = \frac{1}{\ell^{n-1}}  .
\end{split}
\end{equation}
This is sensible, as it suggests that a random unitary operator will (at leading order in $k$ and $\ell$) create maximally mixed reduced states on the radiation and environment factors.
Using the standard formula for the entropy in terms of these traces as described above \eqref{eq:entropy-diff}, we find
\begin{equation}
\begin{split}
    S(\sigma_R) & = \log k , \\
    S(\sigma_\en) & = \log \ell .
\end{split}
\end{equation}
Finally, the coherent information (\ref{eq:coherentinfo}) for a random error of fixed Kraus rank $\ell$ acting on a radiation space of dimension $k$ is
\begin{equation}
    I_c \left( \mathcal{E}, \frac{1}{k}\mathbb{1}_R \right) = S(\sigma_R) - S(\sigma_\en) = \log \frac{k}{\ell} .
\end{equation}
According to this result, a random error must be of sufficient Kraus rank in order to affect the interior.
In particular, utilizing \eqref{CIcond}, if
\begin{equation}
    \log \frac{k}{\ell} \ll -(S_{BH} - \log d_i) 
\end{equation}
then the error $\cE$ on the interior is not quantum correctable based on our criterion on the coherent information.
This result is equivalent to \eqref{eq:thirdcondition}, as advertised.

If the Kraus rank of the randomizing error is scaling with a fractional power of the radiation dimension $k$, for instance $\ell \sim \sqrt{k}$, this shows that a typical random error will actually not affect the interior because the ratio $k/\ell$ will be large, leading to a positive coherent information. 
This is  surprising because after the Page time we have $k \gg e^{S_0}$, and a random error with $\ell = \sqrt{k}$ is modeled by a Stinespring dilation $U_\cE$ that will almost certainly have exponential quantum complexity in $S_0$.
Our result shows that such exponential complexity errors cannot affect the interior.

It also shows that the erasure/SWAP example we discussed previously is somehow finely tuned, because the SWAP Stinespring dilation only requires an environment of dimension $\ell = k$, and for this relation $\log (k/\ell) = 0$.
Of course, for $\ell = k$ we are certainly missing contributions in our Haar integrals as we assumed either $\ell \gg k$ or $k \gg \ell$, so perhaps these terms provide large corrections to the coherent information for a typical error with $\ell = k$ that brings them closer to the SWAP error.
On the other hand, when the Kraus rank of the error is significantly greater than the radiation dimension, it is quite easy to affect the interior after the Page time $k \gg e^{S_0}$, as e.g. we may take $\ell = k^2$ maximal \cite{NielsenChuang} and then we have the condition $\log \frac{1}{k} \ll -S_0$ which is nothing but the statement that we have passed the Page time.

\section{Discussion}\label{sec:disc}
We used a toy model for black hole evaporation in JT gravity to analyze novel quantum error correction properties of the black hole interior. Similar properties  were previously proposed by Kim, Preskill and Tang on the basis of a pseudorandom encoding model for Hawking radiation \cite{Kim:2020cds}, while we investigated these properties in a simple gravitational model.
Modeling errors in the encoded state as  quantum channels $\mathcal{E}$, we found that any error is correctable provided (i) the channel does not have access to black hole microstate details (or more precisely, the encoding map for the code subspace), and (ii) its coherent information $I_c$ with respect to the maximally mixed radiation state is greater than a certain (negative) lower bound set by the black hole entropy. Thus, while the black hole is macroscopic,
its interior
is robustly protected against generic, low-rank errors (i.e. errors whose rank is smaller than the size of the radiation Hilbert space) on the radiation. 
The breakdown of correctability upon violating the coherent information bound happens due to the dominance of the replica wormhole geometry in the gravitational path integral, which is the same effect that leads to restoration of the Page curve in entropy calculations.

\subsection{Robust QEC in the Python's Lunch}
In this work, we have focused on the error correction properties of the black hole interior, viewed past the Page time as being encoded in the radiation bath. It is interesting to ask how the robust error correction properties of the black hole interior that we have found may  generalize to other entanglement wedges. Recent work \cite{Brown:2019rox, Engelhardt:2021mue, Engelhardt:2021qjs} has highlighted the rich structure of entanglement wedges in AdS/CFT. Let $A$ be a subregion, or, more precisely, the domain of dependence $D(A)$ of a subregion $A$, in the boundary CFT and let $\Abar$ be its complement. We can divide the entanglement wedge $\mathcal{W}(A)$ of $A$ in the dual geometry into two portions: (i) the \emph{simple wedge}, i.e., the portion of $\mathcal{W}(A)$ which lies between $A$ and the outermost quantum extremal surface homologous to $A$, and (ii) the \emph{Python's Lunch}, i.e., the complement of the simple wedge inside $\mathcal{W}(A)$. The simple wedge can further be divided into the \emph{causal wedge}, i.e., the portion of the simple wedge which is in causal contact with $D(A)$, and its complement inside the simple wedge. It is expected that the encoding map for semi-classical degrees of freedom in the Python's Lunch portion of an entanglement wedge will be exponentially complex, while the encoding map will not be complex in the simple wedge. Indeed, the black hole interior past the Page time is an example of a Python's Lunch, as it is hidden behind a non-minimal QES, namely the empty surface.

It is natural to conjecture that in more general entanglement wedges, \emph{the encoding of semi-classical degrees of freedom in the Python's Lunch should be robust against generic, low-rank errors on the boundary subregion $A$.}\footnote{This conjecture also makes precise the idea that the holographic encoding of black hole microstates in the AdS/CFT correspondence is exceedingly complex, being essentially random, and is therefore difficult to decode without access to operations that have direct knowledge of the microstates \cite{Balasubramanian:2005mg}.}
By ``generic'', we mean quantum operations which do not have prior access to the details of the microscopic encoding map.\footnote{One way to make this more concrete is by considering typical errors, namely errors whose Stinespring dilation consists of a typical unitary from the Haar ensemble.} In contrast, we do not expect robust quantum error correction in the simple wedge. Indeed, it is not difficult to construct examples of generic, low-rank quantum operations which corrupt the encoding of degrees of freedom in the simple wedge. For instance, in the causal wedge, even a one-qubit environment interacting with $A$ via coupling to some single-trace operator is sufficient. Further, the entire simple wedge can be brought in causal contact with $D(A)$ by a relatively simple unitary operation on $A$. Our conjecture can be analysed in simplified models of holography, such as random tensor networks or fixed area states, as we will report elsewhere \cite{coming_soon}. It would also be interesting to understand the implications of our results in cosmological settings \cite{Langhoff:2021uct,Bousso:2022gth}, de Sitter universes \cite{Shaghoulian:2022fop}, and for disjoint gravitating universes that share entanglement between their degrees of freedom \cite{Balasubramanian:2020coy,Anderson:2020vwi, Balasubramanian:2021wgd}.  

\subsection{Pseudorandomness and complexity}
Our analysis of QEC in the black hole interior has relied entirely on the gravitational path integral, and provides a different perspective on the results of \cite{Kim:2020cds}, where the authors relied on quantum information theory arguments together with an assumption on the pseudorandomness of the radiation density matrix. More precisely, the authors of \cite{Kim:2020cds} assumed that any low-complexity (i.e., polynomial in the black hole entropy) quantum operation could not reliably distinguish the radiation density matrix from the maximally mixed state. This led them to conclude that the black hole interior was robust against low-complexity errors on the bath.  On the other hand, our analysis relied upon a different key assumption: the action of the error channel $\mathcal{E}$ should be independent of gravitational overlaps between black hole microstates $\inner{\psi_{i,i'}^\alpha}{\psi_{j,j'}^\beta}_B$.
If this assumption were violated, our calculation in Sec.~\ref{sec:qec-pssy} would have to be modified because we would potentially need to include gravitational path integrals on additional manifolds in order compute the amplitudes associated to the action of $\cE$.
We are thus led to conclude that the interior is robust against (low-rank) errors which do not have prior access to details of black hole microstates (or more generally, the encoding map for the code). It would be nice to further explore the connection between these two approaches.

It is worth noting that in our analysis, the maximally mixed state on the radiation arises naturally as the state relative to which the coherent information of $\mathcal{E}$ should be computed in order to determine whether or not the error is correctable. This is  an output of the gravitational path integral in the specific toy model we are considering. We can interpret this maximally mixed state loosely as the semi-classical state of the radiation, analogous to Hawking's density matrix, i.e., the coarse-grained, random approximation to the true, pseudorandom density matrix on the radiation.

One curious aspect which deserves further exploration, given this observation, is the following.
In the toy model, the gravitational path integral is really computing an ensemble average $\mathbb{E}[\cdot]$ over many boundary theories \cite{Penington:2019kki}.
The maximally mixed radiation state which appears in our criterion happens to also be the ensemble average of the microscopic radiation states in these theories, because these states depend on gravitational overlaps that will vary in each individual theory in the ensemble. 
So, we may write our criterion for interior state corruption as
\begin{equation}
    I_c(\cE, \mathbb{E}[\rho_R] ) \ll -(S_{BH} - \log d_i ) .
\end{equation}
Though a complete generalization of our criterion to include cases where $\cE$ has access to microstate details was beyond the scope of our work, we may conjecture that a sufficient criterion for interior state corruption that includes cases where $\cE$ has access to microstate details should simply move the ensemble average outside to reflect the fact that $\cE$ also has nontrivial ensemble dependence.
\begin{equation}
    \mathbb{E} [I_c(\cE , \rho_R)] \ll -(S_{BH} - \log d_i) \quad \xRightarrow[]{?} \quad \rho_\inte \text{ is not recoverable from } \cE(\rho_R) .
\label{eq:conjecture}
\end{equation}
Note that in the above conjecture it is crucial that the implication is only to the right.
We are not claiming that the left hand side is a necessary criterion for interior state corruption; we only mean that it may be a sufficient criterion in the presence of a channel $\cE$ which depends on the microstate details.
This contrasts with the criterion \eqref{CIcond} derived in Sec.~\ref{sec:qec-pssy}, which was both necessary and sufficient for interior state corruption due to our no-prior-access assumption.
Indeed, the converse of \eqref{eq:conjecture} can be explicitly violated by a simple channel that couples a Petz operator on the radiation to a single environment qubit.

Understanding \eqref{eq:conjecture} in greater detail could shed light on  properties of the microscopic radiation state, connecting with the work of \cite{Kim:2020cds} and potentially bringing complexity theory into the discussion.
Indeed, a simple counterexample to our no-prior-access-to-black-hole-microstate-overlaps assumption would be a channel which uses the Petz reconstruction of bulk operators on the radiation; however it seems plausible that such channels are exponentially complex \cite{Zhao:2020wgp,Gilyen:2020gmg}, and therefore also violate the assumptions of \cite{Kim:2020cds}.
Analyzing such channels which are allowed to access the microscopic state details in a controlled manner provides an avenue for connecting our assumptions with the pseudorandomness ideas of  \cite{Kim:2020cds}.
In this language, the interesting question to ask is: what properties of the channel $\cE$ with access to microstate details allow it to corrupt the interior while violating the left hand side of \eqref{eq:conjecture}?
It would be good to make a more precise connection between the pseudorandomness assumption of KPT and the assumptions which have gone into the present analysis, perhaps through an analysis of channels which are allowed to access the microscopic state details, as our setup allows us to relax the no-prior-access assumption in a controlled manner, while relaxing the pseudorandom assumption of KPT seems difficult.

\subsection{Implications for semi-classical causality}

We found that an error channel acting on the radiation can affect the black hole interior if it has a sufficiently negative coherent information.   This situation can arise when the Euclidean gravity path integral associated to the error channel is dominated by wormhole contributions.   Some of the expected properties of effective field theories can be challenged by the presence of such wormholes \cite{Geng:2021hlu,Bak:2021qbo}.
We focus here on semi-classical causality in the black hole spacetime, which states that an interior EFT operator $\phi$ should commute with EFT operators $\mathcal{Q}_R$ acting on the radiation:
\begin{equation}
    [\phi, \mathcal{Q}_R] = 0 .
\label{eq:microcausal}
\end{equation}
This expression, much like the notion of bulk radial locality in AdS/CFT \cite{Harlow:2018fse}, ought to hold at least in a code subspace of the physical Hilbert space.
However, after the Page time, the interior is located within the entanglement wedge of $R$, and therefore we may reconstruct logical operators such as $\phi$ using operators which act only on $R$ in the physical Hilbert space.
So, it is not obvious whether microcausality in the sense of \eqref{eq:microcausal} will still be respected, even in the code subspace.
To check this, we must compute the commutator in the physical Hilbert space, using a global representation of $\phi$.
By global representation, we mean a representation of $\phi$ which acts on the physical Hilbert space $\cH_B \otimes \cH_R$.
In place of the radiation operator $\mathcal{Q}_R$, we will use the Kraus operators $E_n$, which model  operations we can perform on the Hawking quanta in the radiation.

Correctability of the set of errors $\cE=\{E_m\}$ has an important implication for this issue, namely  consistency  with the semi-classical causal structure of the black hole spacetime \cite{Kim:2020cds}. Let $\phi$ be a bulk operator in the code subspace localized in the black hole interior, 
\begin{equation}
    \phi = \sum_{k'=1}^{d_e} \sum_{i,j=1}^{d_i} \phi_{ij} \ket{i,k'}\bra{j,k'}_{\inte,\ext} \, .
\end{equation}
This is localized on the interior because we have traced over the exterior factor of the Hilbert space; in other words $\phi$ acts as the identity on $\cH_\ext$.
Recalling that the action of the encoding map is $V\ket{i,i'}_\text{bulk EFT} = \ket{\Psi_{i,i'}}_{B,R}$,  let $\cO$, the ``global'' representation of $\phi$ on the physical Hilbert space $\cH_\ph = \cH_B \otimes \cH_R$, be 
\beq\label{GR}
\cO = V\phi V^{\dagger} = \sum_{k'} \sum_{i,j}\phi_{ij}|\Psi_{i,k'}\rangle\langle \Psi_{j,k'}|_{B,R}.
\eeq
By global representation we simply mean the explicit lift of $\phi$ to an operator $\cO$ on $\cH_\ph$ via conjugation by the encoding map $V$.
The exact correctability of $\cE$ is equivalent to the Knill-Laflamme conditions \eqref{KLcond1}:
\beq
P E_m^{\dagger}E_n P = V( \mathbb{1}^{(\inte)}\otimes B^{(\ext)}_{mn})V^\dagger,\;\;\;\forall m,n,
\label{eq:KLcond2}
\eeq
where the operators $B^{(\ext)}_{mn}$ are analogous to those appearing in \eqref{KLcond1} and  $P = \sum_{i,i'}
\ket{\Psi_{i,i'}}\bra{\Psi_{i,i'}}_{B,R}$ is a projector onto the code subspace. Since $\cO$ is localized on the interior factor on the code subspace, the isometry $V$ by definition obeys $V^\dagger V = \mathbb{1}_{\inte,\ext}$, and the combination $PE_m^\dagger E_n P$ acts as the identity on $\cH_\inte$ after conjugation by $V$ \eqref{eq:KLcond2}.  So we find
\beq
\left[\cO, P E_m^{\dagger}E_n P\right] =0,\;\;\;\forall m,n \, .
\eeq
This is an operator algebra restatement of the error correction conditions \cite{2007PhRvA..76d2303B}. From equation \eqref{GR}, we also observe that $\mathcal{O}$ commutes with the code subspace projector $P= VV^\dagger$ and so this condition  translates to
\beq
P\left[\cO,  E_m^{\dagger}E_n \right] P=0,\;\;\;\forall m,n .
\eeq
Suppose the class of errors $\cE=\{E_m\}$ includes an operator   proportional to the identity, and, if not, extend it with such an operator.  Now require correctability under the extended set \cite{Kim:2020cds}). Then the Knill-Laflamme conditions immediately imply
\beq
P\left[\cO,  E_n \right] P=P\left[\cO,  E_n^{\dagger} \right] P= 0,\;\;\;\forall n,
\label{eq:code-subspace-causality}
\eeq
and so the Kraus operators respect the semi-classical causal structure of the black hole spacetime within the code subspace. 

Thus, any correctable quantum operation $\cE$ on the radiation bath will respect semi-classical causality with respect to operators in the black hole interior. For a correctable quantum operation to violate semi-classical causality, it therefore seems to need prior access to black hole microstate details through the overlaps $\inner{\psi_{i,i'}^\alpha}{\psi_{j,j'}^\beta}$, which would modify the results we obtained in Sec.~\ref{sec:qec-pssy}.
On the other hand,  the failure of correctability due to a large Kraus rank for $\cE$ \emph{does not} necessarily imply a violation of causality, for example if the information is simply moved to another physical environment.
In that case the original representation of $\cO$ on $B\cup R$, the union of of the quantum system dual to the black hole and the radiation bath, fails to reconstruct the bulk operator in the interior.  One must instead construct interior operators on the joint system $B\cup R\cup \en$.

The global representation of $\phi$ need not commute with errors that are uncorrectable, as our derivation above concerning semi-classical causality relied on the Knill-Laflamme conditions for correctability.
A way out of this apparent violation of causality may be that the global representation of $\phi$ is simply inaccessible to any reasonable observer living in the radiation.  In that case, violation of causality for observers in the radiation would not be diagnosed by non-commutation of the Kraus operators with the global representation.   Instead causality would be violated for such observers if the Kraus operators did not commute with a representation of $\phi$ which lives purely on the radiation Hilbert space.
We will now demonstrate that, at least for one particular such representation, semi-classical causality is indeed never violated.

An explicit example of such a representation on $R$ alone is the Petz reconstruction
\beq
\cO_\text{Petz} = \cE(P)^{-1/2}\cE(V\phi V^{\dagger})\,\cE(P)^{1/2} .
\eeq
This operator makes explicit use of the quantum error channel acting exclusively on the radiation.  If there is no recovery channel, i.e., if the error cannot be corrected, then the Petz operator will vanish \cite{Penington:2019kki}.
This is the case, for example, before the Page time.
Thus, this operator respects causality, because it vanishes precisely when the channel is not correctable.  It is possible for this to happen because operator makes explicit use of the known error channel. When the error is correctable, the Petz operator has code subspace matrix elements which match the global reconstruction, and so causality is ensured in this case as well, because by (\ref{eq:code-subspace-causality}) the Kraus operators will also commute with $\cO_{\text{Petz}}$ in the code subspace.
When the error channel is correctable, $\phi$ and $\cO_{\text{Petz}}$ are identical as operators acting on the code subspace.
The global representation, on the other hand, is always nonzero on the physical Hilbert space if the bulk EFT operator $\phi$ is nonzero, and the Knill-Laflamme conditions are not enough to ensure causality for such operators.
The bottom line is that an operator represented purely on the radiation with access to the error channel, such as the Petz operator, may be constructed to vanish when the error may violate semi-classical causality with respect to the global representation, thus ensuring that this weaker version of semi-classical causality is preserved.

\subsection*{Acknowledgments}

We thank Abhijit Gadde, Lampros Lamprou, Raghu Mahajan, Gautam Mandal, Alexey Milekhin, Shiraz Minwalla, Daniel Ranard, Pratik Rath, Pushkal Shrivastava, Douglas Stanford, Sandip Trivedi and Zhenbin Yang for useful discussions. We thank Abhijit Gadde for comments on a previous version of the paper. 
VB is supported in part by the Department of Energy through grant DE-SC0013528 and grant QuantISED DE-SC0020360, as well as the Simons Foundation through the It From Qubit Collaboration (Grant No. 38559). 
AK is supported by the Simons Foundation through the It from Qubit Collaboration.
CL is supported by the Department of Energy through grants DE-SC0013528 and DE-SC0020360. OP is supported by the Department of Atomic Energy, Government of India, under project identification number RTI 4002.

\appendix

\section{Haar integration over unitary matrices}
\label{sec:haar}

Here we provide details of the evaluation of Haar integrals in Sec.~\ref{sec:randomerrors}.
As we are most concerned with integrals over $U(k\ell)$ where $1 \ll k \ll \ell$ or $1 \ll \ell \ll k$, there is a competition between the natural $1/(k\ell)$ suppression of the so-called Weingarten functions and particular Kronecker structures which can act to cancel some of the summations.\footnote{A similar analysis appears in a somewhat different context in \cite{Balasubramanian:2021mxo}.  See there for several more detailed examples. }
The general formula for the Haar integration in terms of these functions and Kronecker deltas is
\begin{equation}\label{HaarInt}
    \int_{U(D)} U^{i_1}_{j_1} (U^\dagger)^{j_1'}_{i_1'} \cdots U^{i_n}_{j_n} (U^\dagger)^{j_n'}_{i_n'} \; dU = \sum_{\mu,\tau \in S_n} \delta^{i_1}_{i'_{\mu(1)}} \delta^{j_1}_{j'_{\tau(1)}} \cdots \delta^{i_n}_{i'_{\mu(n)}} \delta^{j_n}_{j'_{\tau(n)}} \times \text{Wg}(\mu \tau^{-1}, D) ,
\end{equation}
where $\mu$ and $\tau$ are elements of the permutation group $S_n$ and the asymptotic behavior of the Weingarten function Wg$(\tau, D)$ for $D \gg 1$ is 
\begin{equation}
    \text{Wg}(\tau \in S_n,D) \sim D^{-n-|\tau|} \prod_{i \in \text{cycles}} (-1)^{|a_i|-1} C_{|a_i|-1},
    \label{eq: Weingarden}
\end{equation}
where $|\tau|$ is the minimal number of transpositions\footnote{A transposition is a permutation which swaps two elements and leaves the others fixed.} needed to form $\tau$, $a_i$ is the length of the $i^{\text{th}}$ cycle in $\tau$, and $C_n = (2n)!/(n!(n+1)!)$ is the $n^{\text{th}}$ Catalan number.
The large $D$ limit is relevant here because, in our case, $D = k\ell$, and we are considering a large radiation Hilbert space of dimension $k$ and errors with Kraus rank $\ell$.
 
\subsection{$f$-Tensor}
 
In Sec.~\ref{sec:f-tensor-haar}, we needed to average the $f$-tensor over the unitary group in the Haar measure.  We seek to compute the following integral:
\begin{align}
\begin{split}
&\int_{U(D)} dU  f_{\alpha_1,\beta_1,\cdots \alpha_n,\beta_n}(U) \\
& = \sum_{m_1,\cdots, m_n}\sum_{\gamma_1,\cdots, \gamma_n} \int_{U(D)} dU  (U^{\dagger})^{\alpha_1,e_0}_{\gamma_1,e_{m_2}}  U^{\gamma_1,e_{m_1}}_{\beta_1,e_0}  \cdots (U^{\dagger})^{\alpha_n,e_0}_{\gamma_n, e_{m_1}}   U^{\gamma_n,e_{m_n}}_{\beta_n,e_0} \\
& = \sum_{m_1,\cdots, m_n}\sum_{\gamma_1,\cdots, \gamma_n} \sum_{\mu ,\tau \in S_n} \delta^{\gamma_1 m_1}_{\mu(\gamma_1 m_2)} \delta^{\tau(\alpha_1)}_{\beta_1} \cdots \delta^{\gamma_n m_n}_{\mu(\gamma_n m_1)} \delta^{\tau(\alpha_n)}_{\beta_n} \text{Wg}(\mu \tau^{-1}, k\ell)\\
& = \sum_{m_1,\cdots, m_n} \sum_{\gamma_1,\cdots, \gamma_n} (\delta^{\gamma_1 m_1}_{\gamma_1 m_2} \cdots \delta^{\gamma_n m_n}_{\gamma_n m_1} \delta^{\alpha_1}_{\beta_1} \cdots \delta^{\alpha_n}_{\beta_n} +\delta^{\gamma_1 m_1}_{\gamma_n m_1} \cdots \delta^{\gamma_n m_n}_{\gamma_{n-1} m_n} \delta^{\alpha_n}_{\beta_1} \cdots \delta^{\alpha_{n-1}}_{\beta_n} ) \times \text{Wg}(1^n, k\ell) + \cdots \\
& = \frac{1}{(k\ell)^n} \left( k^n \ell (\delta^{\alpha_1}_{\beta_1} \cdots  \delta^{\alpha_n}_{\beta_n}) + (k \ell^n ) (\delta^{\alpha_n}_{\beta_1} \cdots \delta^{\alpha_{n-1}}_{\beta_n} ) \right)   + \cdots\\
& = \ell^{1-n} (\delta^{\alpha_1}_{\beta_1} \cdots  \delta^{\alpha_n}_{\beta_n}) + k^{1-n} (\delta^{\alpha_n}_{\beta_1} \cdots \delta^{\alpha_{n-1}}_{\beta_n} )   + \cdots
\label{eq: f-tensor-complete}
\end{split}
\end{align}
where in the second to last line the first and second terms have $k$ factors of the radiation indices $\{\gamma_i\}$ contracted, the second term has $\ell$ factors of $\{m_i\}$ contracted, and in the last line we used the Weingarten function defined in \eqref{eq: Weingarden}. The ellipsis includes factors that have larger Weingarten suppression with $\mu\neq \tau$ and the tensor structure that correspond to non-replica-symmetric geometries, since these tensors will contract with the tensor structure that comes from the bulk path integral in Sec.~\ref{sec:randomerrors}. To see the Weingarten suppression, consider $\mu \tau^{-1} = \sigma \in S_n$. Decomposing $\sigma$ into $g$ transpositions, one observes that $\text{Wg}(\sigma, k\ell) \sim (k\ell)^{-n} (-2) \left(\frac{-2}{k\ell}\right)^{g}$, which will be suppressed exponentially by the difference between $\mu$ and $\tau$ when $k\ell>2$.
Therefore, for a fixed geometry, the largest contribution to the the entropy comes from setting $\mu=\tau$. 

As observed in \ref{sec:f-tensor-haar}, a tensor $\delta^{\alpha_j}_{\beta_j}$ in the Haar integration that contract with a given bulk geometry represented by the index structure $\delta^{\alpha_j}_{\beta_j}$ Haar integration reinforces different gravity configurations, in the same manner as \eqref{eq: g-f-contraction1} and \eqref{eq: g-f-contraction2}, because $\sum_{\alpha_j,\beta_j} (\delta^{\alpha_j}_{\beta_j})^2 = k$ where $k$ is the number of $\alpha$- and $\beta$-components. Consider an element $\tau \in S_n$ that a geometry that is not replica symmetric, i.e. it gives a $\delta^{\alpha}_{\beta}$ structure that does coincide with either \eqref{eq:DiscOverlap} or \eqref{eq:conn-EOW-structure}, e.g. $\tau=(n'\;1\;2\; \cdots n'-1)$ a $m$-cycle shorter than $n$. As shown above, the term that has minimal Weingarten suppression given this geometry has $\mu = \tau$. However, the choice of $\mu$ results in $n'$ matched $\gamma$-indices and $(n'-n)$ mismatched $m$-indices in the third line of \eqref{eq: f-tensor-complete}, contributing $k^{n'-n}\ell^{n'}$ to this tensor $\delta^{\beta_1}_{\alpha_n'}\cdots \delta ^{\beta_n'}_{\alpha_{n'-1}}\delta^{\beta_{n'+1}}_{\alpha_{n'+1}}\cdots\delta^{\beta_n}_{\alpha_n}$. From this one can see that with the same value of the Weingarten function, the dominating contribution to the entropies is either the completely disconnected or the connected cases in either limit $k \gg \ell$ or $k \ll \ell$ respectively.   This observation also agrees with the two limits in \cite{Penington:2019kki} while they sum over all the planar geometries. Tempting as it is to think that every element in the sum of the symmetric group will reinforce a type of geometry, it is more natural to consider the cyclic group instead, given the boundaries lie on a circle. Unlike \cite{Penington:2019kki}, it is not clear that the planar geometries has larger weights than the geometries with crossings directly from the Haar integration. However, one can conjecture that a typical error on the radiation can somehow control the entropy of the states in the bulk through different wormholes enabled by replica theory.

\subsection{Coherent information criterion}

In Sec.~\ref{sec:CIcond-haar}, we are interested in the Haar average of the following R\'enyi entropies:
\begin{equation}
\begin{split}
    \Tr \sigma_R^n & = \frac{1}{k^n} \sum_{\alpha_1, \cdots , \alpha_n} \sum_{\beta_1, \cdots , \beta_n} \sum_{m_1, \cdots , m_n} U^{\beta_1,e_{m_1}}_{\alpha_1,e_0} (U^\dagger)^{\alpha_1,e_0}_{\beta_2, e_{m_1}} \cdots U^{\beta_n,e_{m_n}}_{\alpha_n,e_0} (U^\dagger)^{\alpha_n,e_0}_{\beta_1, e_{m_n}} , \\
    \Tr \sigma_\en^n & = \frac{1}{k^n} \sum_{\alpha_1, \cdots , \alpha_n} \sum_{\beta_1,\cdots \beta_n} \sum_{m_1, \cdots , m_n} U^{\beta_1,e_{m_1}}_{\alpha_1,e_0} (U^\dagger)^{\alpha_1,e_0}_{\beta_1,e_{m_2}} \cdots U^{\beta_n,e_{m_n}}_{\alpha_n,e_0} (U^\dagger)^{\alpha_n,e_0}_{\beta_n,e_{m_1}} .
\end{split}
\end{equation}
In terms of the permutation group element sum in \eqref{HaarInt}, the only relevant contribution comes from the ``free'' configuration $\mu = \tau = e$, as this has minimal Weingarten suppression and collapses only a single sum (by setting one pair of indices to be equal via a delta function) in both R\'enyi entropies.
There is just one other configuration which also collapses only a single sum, but which has maximal Weingarten suppression ($\mu$ is an $n$-cycle and $\tau = e$ is the identity), so we discard it. 
Thus, all configurations other than the free one are suppressed in either $1/k$ or $1/\ell$.
This choice results in maximally mixed R\'enyi entropies, as expected for the leading behavior of a random unitary Stinespring dilation.

\bibliographystyle{JHEP}
\bibliography{refs}

\end{document}